\documentclass[a4paper,11pt]{article}

\usepackage{jheppub} 

\usepackage[T1]{fontenc}
\usepackage{epsfig}
\usepackage{latexsym}
\usepackage{amsfonts}
\usepackage{amsmath}
\usepackage{amsthm}
\usepackage{amssymb}
\usepackage{amsbsy}
\usepackage{multirow}
\usepackage{mathrsfs}
\usepackage{textcomp}
\usepackage[vcentermath, enableskew]{youngtab}

\def\calb         {{\cal B}}

\def\cale         {{\cal E}}

\def\cali         {{\mathscr I}}
\def\calj         {{\cal J}}
\def\calk         {{\cal M}_0}
\def\call         {{\cal L}}
\def\calm         {{\cal M}}
\def\caln         {{\cal N}}
\def\calo         {{\cal O}}
\def\calp         {{\cal P}}

\def\cals         {{\cal S}}
\def\calt         {{\cal T}}

\def\calv         {{\cal V}}

\def\mbbc         {\mathbb{C}}

\def\mm{{\cal V}}

\def\mE{{\cale}}

\def\eulnum{{\mathscr E}}

\def\a{\alpha}
\def\b{\beta}

\def\g{\gamma}

\def\d{\delta}

\def\l{\lambda}

\def\lra{\longrightarrow}
\def\Lra{\Longrightarrow}

\def\k{\kappa}

\def\m{\mu}
\def\mCP{\mathbb{CP}}
\def\mGr{\mathbb{G}\mathrm{r}}
\def\mbbc{\mathbb{C}}

\def\mbi{\mathbb{I}}

\def\mbb{\mathbb{B}}

\def\O{\Omega}

\def\q{\theta}

\def\ttd{\mathrm{\bf{td}}}
\def\ttc{\mathrm{\bf{tc}}}
\def\tch{\mathrm{\bf{ch}}}

\def\x{\xi}

\def\Gr{{\mathbb{G}\mathrm{r}}}

\title{\boldmath Pure-Higgs states from the Lefschetz-Sommese theorem}

\author[a]{I. Messamah}
\author[b,c]{and D. Van den Bleeken}

\affiliation[a]{Laboratoire de Physique des Particules et Physique Statistique, \\
Ecole Normale Sup\'{e}rieure-Kouba, BP92, \\
Vieux-Kouba, 16050 / Algiers, Algeria}
\affiliation[b]{Physics Department, Bo\u{g}azi\c{c}i University, \\34342 Bebek / Istanbul, Turkey}
\affiliation[c]{Secondary address: Institute for Theoretical Physics,\\ KU Leuven 3001 Leuven, Belgium}
\emailAdd{ilies.messamah@g.ens-kouba.dz}
\emailAdd{dieter.vand@boun.edu.tr}

\abstract{We consider a special class of N=4 quiver quantum mechanics relevant in the description of BPS states of D4D0 branes in type II Calabi-Yau compactifications and the corresponding 4-dimensional black holes. These quivers have two abelian nodes in addition to an arbitrary number of non-abelian nodes and satisfy some simple but stringent conditions on the set of arrows, in particular closed oriented loops are always present. The Higgs branch can be described as the vanishing locus of a section of a vector bundle over a product of a projective space with a number of Grassmannians. The Lefschetz-Sommese theorem then allows to separate induced from intrinsic cohomology which leads to the notion of pure-Higgs states. We compute explicit formulae for an index counting these pure-Higgs states and prove -- for this special class of quivers -- some previously stated conjectures about them.}

\begin{document} 
\maketitle
\flushbottom

\section{Introduction and overview}
Quiver quantum mechanics \cite{Denef:2002ru} provides a powerful effective description of BPS states in $\caln=2$ gauge theory and type II string/supergravity theory, see for example \cite{Cecotti:2013sza, Pioline:2013wta} for reviews. The BPS states correspond to the supersymmetric groundstates of the quiver theory and depending on the coupling can be described geometrically in terms of the Coulomb or Higgs branch \cite{Denef:2002ru}. Recently various powerful techniques have been developed in the study of these supersymmetric groundstates, such as the Coulomb branch formula \cite{Manschot:2014fua, Beaujard:2019pkn}, localization \cite{Hori:2014tda, Ohta:2014ria, Cordova:2014oxa}, mutations \cite{Manschot:2013dua, Kim:2015fba} and others \cite{Lee:2013yka, Alexandrov:2018iao}. States on the Coulomb branch can be matched rather directly to a gravitational description, since both pictures describe spatial bound states of dyonic centers \cite{Denef:2000nb, Bates:2003vx, Denef:2002ru}. The Higgs branch description, where the groundstates are represented as cohomology classes, appears more fundamental however, since  it contains not only the Coulomb branch ground states but in certain cases a large number of additional ground states as well \cite{Denef:2007vg, Bena:2012hf, Lee:2012naa, Lee:2012sc}. These extra states that do not have a Coulomb branch description were named pure- or intrinsic Higgs states. Their presence is related to at least one closed loop in the quiver diagram, which is equivalent to the existence of superpotential, although this is not a sufficient condition. The existence of such states raises the interesting question of how they should be identified in the supergravity regime\footnote{Due to supersymmetry and the existence of a protected index they cannot simply disappear in the gravitational regime.}. It has been proposed that the pure-Higgs states are exactly those states that form the microscopic description of single centered extremal black holes. This is because they are, at least in all examples studied, angular momentum singlets\footnote{See e.g. \cite{Chowdhury:2014yca, Chowdhury:2015gbk, Heidmann:2018vky} for arguments that the microstates of BPS black holes have zero angular momentum.} that are stable under all wall-crossing. This is a unique combination of features that in N=2 supergravity is shared only by single center black holes. Furthermore the index counting pure-Higgs states, a quiver invariant, coincides with the single center black hole contributions to the Coulomb branch formula \cite{Manschot:2012rx, Manschot:2013sya} and additionally it is known that the number of pure-Higgs states grows exponentially in the charges \cite{Denef:2007vg, Bena:2012hf} leading to a non-negligible macroscopic entropy. Still, a precise derivation of the Bekenstein-Hawking entropy of a black hole by the counting of pure-Higgs states -- presumably among multiple quivers -- remains lacking. 

We can summarize the conjectured properties of pure-Higgs states -- defined to be those states on the Higgs branch that have no counterpart on the Coulomb branch -- more concretely as follows:
\paragraph{Pure-Higgs conjectures}
\begin{itemize}
	\item[1)] pure-Higgs states are singlets under the Lefschetz SU(2)
	\item[2)] the number of pure-Higgs states is independent of the FI parameters of the quiver
	\item[3)] pure-Higgs states exist only if the Coulomb branch supports at least one angular momentum singlet
\end{itemize}
The first conjecture was already mentioned above, with the understanding that the Lefschetz SU(2) action on the cohomology of the Higgs branch translates as the action of angular momentum SU(2) on space-time. Phrased as in 1) this is equivalent to the pure-Higgs states all being located in the middle cohomology. The 2nd conjecture translates as the stability under wall-crossing. The third conjecture can be re-phrased as the existence of pure-Higgs states being related to the existence of so called 'scaling solutions' \cite{Denef:2007vg} to the Denef equations \cite{Denef:2000nb} describing the Coulomb branch/supergravity solutions. The phrasing as in 2) is a bit more precise since it takes into account some quantum corrections to the angular momentum.

In most previous examples \cite{Denef:2007vg, Bena:2012hf, Lee:2012naa, Lee:2012sc, Manschot:2012rx}, based on quivers with only abelian nodes, the Higgs branches supporting pure-Higgs states could be identified with complete intersection manifolds. By the Lefschetz hyperplane theorem it follows that the cohomology of such spaces splits into an induced part, inherited from the ambient space, and an intrinsic part particular to the intersection. By direct observation the induced cohomology can be matched onto the Coulomb branch spectrum, and this in turn identifies the pure-Higgs states with the intrinsic cohomology. Property 1) is then a direct consequence of the Lefschetz theorem. Properties 2) and 3) can then be verified by direct computation as well. 

In this paper we will explore pure-Higgs states in a new class of examples, where the Higgs branch is the vanishing locus of a section of a vector bundle, a notion generalizing that of complete intersection. Interestingly there exists a generalization of the Lefschetz hyperplane theorem to this setting, known as the Lefschetz-Sommese theorem \cite{Lazarsfeld-R}.

\begin{figure}[tbp]
	\centering
	\includegraphics[scale=1]{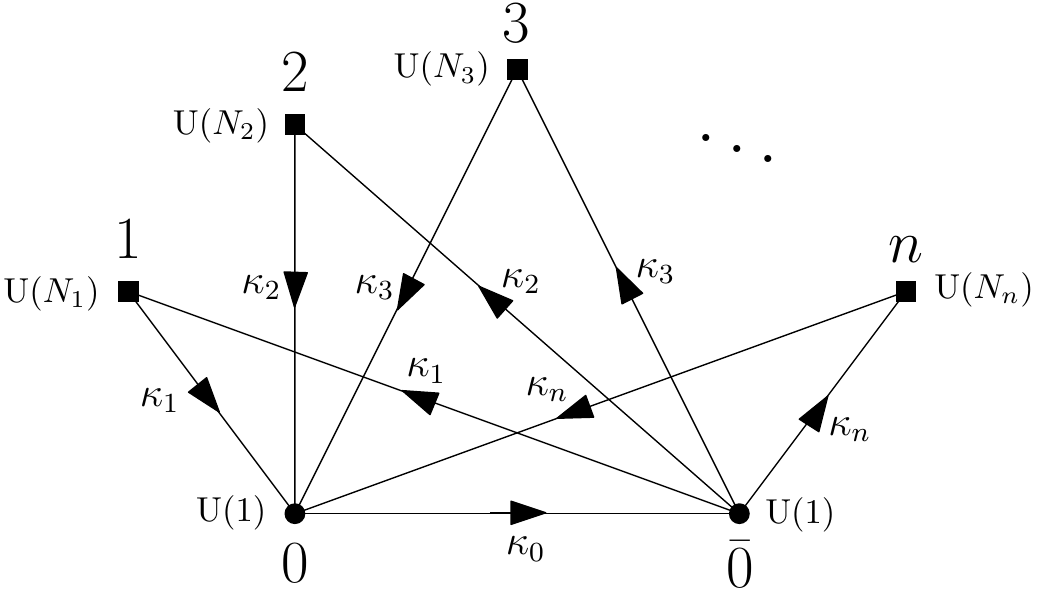}
	\caption{The quiver diagram of a $2+n$ node quiver.}\label{2nquiv}
\end{figure}

We'll refer to the class of quivers we study as $2+n$ node quivers, since they have two nodes with an abelian U(1) gauge group and $n$ nodes with non-abelian gauge group, U($N_i$) respectively. We allow an arbitrary number, $\kappa_0$, of arrows between the abelian nodes. There are no arrows between pairs of non-abelian nodes, but each of the non-abelian nodes is connected in a symmetric fashion to the two abelian nodes, by $\kappa_i$ arrows respectively, oriented in such a way that a separate loop is formed along the 2 abelian nodes and each one of the non-abelian nodes\footnote{Other quivers with oriented loops and non-abelian nodes have been considered in \cite{Manschot:2012rx, Lee:2013yka, Kim:2015fba, Beaujard:2019pkn}, except for the single example in section 3.2.2 of \cite{Lee:2013yka} they all have a different arrow structure than the quivers we study. The quivers discussed in section 6.1 and 6.2 of \cite{Manschot:2012rx} match to a 2+1 node quiver in the limit $b\rightarrow c$. Although this limit violates the assumption $b<c$ made in \cite{Manschot:2012rx} it correctly reproduces the Coulomb branch/induced cohomology part of the result we find in this work.}. See figure \ref{2nquiv} for the diagram representing a $2+n$ node quiver.

Our main motivation to study $2+n$ node quivers is that they are (a slight generalization of) the quivers describing the bound states of fluxed D6 and anti-D6 branes together with an arbitrary number of D0 branes \cite{Denef:2007vg} that appear in type IIA Calabi-Yau compactifications and that carry the same total charge as the D4D0 black hole\footnote{See e.g. \cite{Bena:2017geu} for a connection to the D1D5 system and the superstrata program \cite{Shigemori:2020yuo}.} \cite{Maldacena:1997de}. It was shown in \cite{deBoer:2009un} that the Coulomb branch states of these quivers match the perturbative supergravity spectrum but fall short in reproducing the entropy of this black hole and it is thus an interesting question to ask if the pure-Higgs states could fill the gap\footnote{Footnote a similar gap has recently been observed in \cite{Martinec:2019wzw} in a closely related context.}. The results of this paper are a first step towards answering that question, but we leave an investigation of the growth of states for asymptotic values of the parameters for future work.

The D-term equations of the 2+$n$ quivers define an ambient space that is the product of a projective space with a number of Grassmannians, to be precise $\calm=\mathbb{CP}^{\kappa_0-1}\times \prod_i \Gr (N_i,\kappa_i)$, at least for particular values of the FI parameters. The superpotential made possible by the oriented loops leads to F-term conditions that have the mathematical interpretation of a section of a complex vector bundle over the ambient manifold.  The Lefschetz-Sommese theorem splits the cohomology into induced and intrinsic parts, with the intrinsic classes restricted to the middle cohomology. It is thus suggestive to identify also in this setting the intrinsic cohomology with the pure-Higgs states. Using detailed knowledge of the cohomology of the ambient space we can verify this for the full class of examples, by matching the induced cohomology to the Coulomb branch states, which then implies 1). We then derive an explicit formula for an index counting the pure-Higgs states and then check it satisfies 2) and 3). To do so we compute the holomorphic Euler number of the Higgs branch manifold -- or more precisely a refined Witten index -- via the Hirzebruch-Riemann-Roch theorem and techniques from algebraic topology\footnote{The be precise, in one of the chambers that we consider we face a technical problem in the computation of the holomorphic Euler number via methods of algebraic topology, there we resort to the Abelianization technique of \cite{Lee:2013yka}.}. Interestingly in the  derivation of 2) and 3) only the $\mathbb{CP}$ factor seems to play a role, as the argument remains unchanged if we replace the product of Grassmannians with an arbitrary compact K\"ahler manifold of diagonal cohomology (see \eqref{cohas}). Although we give a precise formula for the pure-Higgs index it remains computationally challenging to evaluate it for large values of the parameters. We provide a list of results for small values of the parameters in table \ref{pHtable}. This table reveals an interesting symmetry that we show has its origin in the Grassmannian factors and we prove it exists for the full class of $2+n$ node quivers. 

The paper is structured as follows. In section \ref{LSsec} we start with a discussion of some consequences of the Lefschetz-Sommese theorem for some geometries that are of relevance. This is a purely topological/geometrical discussion and only serves to introduce some concepts, notation and technical results for later use. In section \ref{sect-3nodes-hggs-branch} we introduce the $2+n$ node quivers, compute the supersymmetric groundstates on their Coulomb and Higgs branch. From these we investigate the pure-Higgs states and using the results of section \ref{LSsec} prove conjectures 1), 2) and 3). In the appendix we collect a few additional details on the geometry and topology of complex Grassmannians.

It is possible to skip section \ref{LSsec} and directly start reading section \ref{sect-3nodes-hggs-branch}, referring back to section \ref{LSsec} when needed.

\section{The Lefschetz-Sommese theorem and applications}\label{LSsec}
The BPS states of interest can be identified with Dolbeault cohomology classes on the Higgs branch $\mm$ associated to a certain quiver quantum mechanics that we introduce in section \ref{sect-3nodes-hggs-branch}. In the special class of quivers we study in this paper $\calv$ has a particular geometrical interpretation that gives some extra tools to study its cohomology, which will be the subject of this section.
\subsection{Generalities}
The geometric interpretation mentioned above is that $\calv$ is the vanishing locus of a section of a complex vector bundle $\cale$ over a compact K\"ahler manifold $\calm$. More colloquially this means that $\calv$ is a subset of $\calm$, formed by the solutions to a number of equations defined on $\calm$.  Since we will be associating the notation to this interpretation it can be useful for future reference to clearly list this once more:
\begin{align}
\calm :&\qquad\mbox{compact K\"ahler manifold}\\ 
\cale :&\qquad\mbox{complex vector bundle over $\calm$}\\
\calv :&\qquad\mbox{vanishing locus of a section of $\cale$}
\end{align}
We will also use the following shorthands in the rest of this paper:
\begin{equation}
 d=\dim_\mathbb{C}\calm \qquad r=\mathrm{rank}_\mathbb{C}\,\cale
\end{equation}
Note that this implies that $\dim_\mathbb{C}\calv=d-r$. If  $\cale$ is ample, which we'll assume in this paper, the Lefschetz-Sommese theorem, see for example \cite[theorem 7.1.1]{Lazarsfeld-R}, almost completely determines the cohomology of $\calv$ in terms of that of the ambient space $\calm$. More precisely

\begin{quote}{\bf Lefschetz-Sommese} : \label{Lefschetz-thm-gen}
	Under the assumptions above there is the following relation between Hodge numbers:
	\begin{align}
	h^{(p,q)} (\mm) = \left\{ \begin{array}{ll}
	h^{(p,q)} \left( \calm \right) & \quad \mathrm{if} \;\;\; (p+q) < d-r\, \;,\\
	h^{(p,q)} \left( \calm \right) + \b^{(p,q)}(\mm) & \quad \mathrm{if} \;\;\; (p+q) =d-r \;, 
	\end{array}
	\right.  
	\end{align}
	where $\b^{(p,q)}(\mm)$ are some non-negative integers. The remaining Hodge numbers of $\calv$ are determined by those above via the relations $h^{(p,q)} (\mm)=	h^{(q,p)} (\mm)$ and $	h^{(p,q)} (\mm)=	h^{(d-r-p,d-r-q)} (\mm)$.
\end{quote}
Restated in words, all the cohomology of $\mm$ originates directly from that of the ambient manifold $\calm$, except for the cohomology of middle dimension, where additional non-trivial elements can be present. We will refer to the part of the cohomology of $\mm$ that originates in that of $\calm$ as the {\it induced cohomology}, while we'll refer to the extra part as the {\it intrinsic cohomology}. The intrinsic cohomology can be identified with the pure-Higgs states in the physical context. The Lefschetz-Sommese theorem  does not determine the amount of intrinsic cohomology, i.e. it leaves the $\b^{(p,q)}$ indeterminate apart from being non-negative integers.

The relevant information on the cohomology can be packaged in an index/generating function. For an arbitrary compact K\"ahler manifold ${\cal K}$ we define 
\begin{equation}
\cali({\cal K} ; t)=t^{-\dim_\mathbb{C}{\cal K}}\sum_{p,q=0}^{\dim_\mathbb{C}{\cal K}}(-1)^{p+q}t^{2p}h^{(p,q)}({\cal K})\label{refinddef}
\end{equation}
We'll refer to \eqref{refinddef} as the {\it refined index}, since it corresponds to the refined Witten index of a quiver quantum mechanics in case ${\cal K}$ is a quiver moduli space, in that setting it also goes under the name of $\chi$-genus \cite{Lee:2012naa, Beaujard:2019pkn}. From a more general geometric perspective the refined index is closely related to the holomorphic Euler number $\eulnum$: 
\begin{equation}
\cali({\cal K} ; t)=t^{-\dim_\mathbb{C}{\cal K}}\eulnum({\cal K};t^2)
\end{equation}
This relation is also of practical use to us, since it allows to re-express the refined index through the Hirzebruch-Riemann-Roch theorem as:
\begin{equation}
\cali ({\cal K}\;;\;t ) = t^{-\dim_\mathbb{C}{\cal K}}\int_{{\cal K}} \ttc \left(\calt{\cal K};t^2\right) \;, \label{rrh-index-thm} 
\end{equation}
here $\calt{\cal K}$ is the holomorphic tangent bundle, while we use the shorthand $\ttc(\calb;t)$ for the product of the Todd class of a bundle $\calb$ with the Chern character of its dual:
\begin{equation}
\ttc(\calb;y)=\ttd(\calb)\cdot\tch(\calb^*;-y)\qquad   \label{new-mult-class} 
\end{equation}

The Lefschetz-Sommese theorem strongly restricts the form of $\cali(\mm)$; the split of the cohomology in an induced and intrinsic part respectively leads to a split of the refined index as
\begin{equation}
\cali(\mm \;;\; t)={\mathscr B}(\mm;t)\, + \cali_\mathrm{ind} (\mm \;;\; t)\label{refindsplit}
\end{equation}
The contribution from the intrinsic part is simply
\begin{equation}
{\mathscr B}(\mm;t)=(-1)^{d-r}\sum_{p=0}^{d-r} t^{2p-d+r} \beta^{(p,d-r-p)}\,,\label{bdef}
\end{equation}
while the induced part  $\cali_\mathrm{ind} (\mm )$ is fully determined in terms of the cohomology of the ambient space $\calm$. Although our main interest lies with ${\mathscr B}(\mm)$, since it corresponds to the quiver invariant \cite{Lee:2012naa} or single center index \cite{Manschot:2014fua} counting pure-Higgs states, there are -- as far as we are aware -- currently no mathematical tools available to 'directly' compute ${\mathscr B}(\mm)$. The 'indirect' route we use is to separately compute $\cali(\mm)$ and $\cali_\mathrm{ind}(\mm)$ and then take the difference to obtain ${\mathscr B}(\mm)$. We'll discuss some more details on both contributions in turn. Before we do that, let us remind the reader that the BPS states of physical interest, and hence also the corresponding cohomology classes discussed in this section, can be organized in SU(2) multiplets. From a physics point of view this is nothing but a classification by angular momentum, while from the geometric perspective this corresponds to the Lefschetz SU(2) \footnote{In the mathematics literature one typically considers the complexified Lie group SL(2), rather than its maximally compact real subgroup SU(2).} action on the cohomology of the K\"ahler manifold ${\cal K}$. It is generated by the K\"ahler form $\calj$ as follows
\begin{align}
J_+ = \calj \wedge \;, \qquad J_- = \iota_\calj \;, \qquad J_3 = \frac{1}{2} \, \left( \mathrm{deg} - \mathrm{dim}_\mathbb{C}{\cal K}\right) \;, \label{sl2-gen}
\end{align}
where $\mathrm{deg}$ is the degree of the form on which $J_3$ acts. Since the induced cohomology on its own forms an SU(2) representation, and since the intrinsic cohomology is all of middle degree, it follows that the intrinsic cohomology of $\mm$ is made out of SU(2) singlets. The decomposition into irreducible SU(2) representations of the induced cohomology is of physical interest, since it provides a map \cite{Bena:2012hf} from the abstract Higgs branch into the Coulomb branch/supergravity. As we'll review next, $\cali_\mathrm{ind}(\mm,t)$ contains all information on the SU(2) representation content, at least for the special class of K\"ahler manifolds that we will consider. 

\subsubsection{The ambient and induced index}
In all examples we will consider, the cohomology of the ambient space $\calm$ is quite simple, in that all its non-trivial cohomology is of even degree, and furthermore diagonal in $p$ and $q$, i.e.
\begin{equation}
h^{(p,q)}(\calm)=b^{(p+q)}(\calm)\, \delta^{p,q}\label{cohas}\,
\end{equation}
with $b^{(i)}$ the Betti numbers. Under this condition the refined index \eqref{refinddef} simplifies and becomes proportional to the Poincar\'e polynomial:
\begin{equation}
\cali ({\calm} ; t) =\; \sum_{i=0}^{2d} b^{(i)} ({\calm}) \; t^{i-d} \;. \label{spin-funct-def}
\end{equation}
Since the power of $t$ is the eigenvalue of the Lefschetz $J_3$, the refined index $\cali ({\calm} ; t)$ is nothing but the character of the SU(2) representation formed by the cohomology of $\calm$. This implies there is the decomposition
\begin{equation}
\cali (\calm \;;\; t) = \sum_{i=0}^{d} m_i (\calm) \; \chi_{i} (t)\;, \label{kahler-spin-funct-spin}
\end{equation}
in terms of irreducible SU(2) characters\footnote{Note that $n$ is a postive integer, equaling twice the physical spin, i.e. $n=2j$.} 
\begin{equation}
\chi_n (t) = \sum_{i=0}^{n} t^{2\,i-n} = \frac{t^{-n-1}-t^{n+1}}{t^{-1}-t}\;. \label{spin-j-function}
\end{equation}
Given the explicit action \eqref{sl2-gen}, the multiplicities $m_i$ are found to be\footnote{We use the convention that $b^{-2} (\calm) = b^{-1} (\calm) = 0$.}
\begin{align}
m_{i} (\calm) = b^{d-i} (\calm) - b^{d-i-2} (\calm) \;, \quad 0 \leq i \leq d \;, \label{poicare-polynom-spin-coeff-1}
\end{align}

This structure in the refined index of the ambient manifold $\calm$ carries through to the induced part of the refined index of the vanishing locus $\calv$, due to the Lefschetz-Sommese theorem \eqref{Lefschetz-thm-gen}. One finds that
\begin{equation}
\cali_\mathrm{ind} (\mm \;;\; t) = \sum_{i=0}^{d-r} m_{r+i} (\calm) \; \chi_{i} (t) \;,\qquad \label{Lefschetz-act-spin-funct-spin}
\end{equation}
By comparing \eqref{Lefschetz-act-spin-funct-spin} to \eqref{kahler-spin-funct-spin} we can conclude that the Lefschetz-Sommese theorem has as effect the removal of all the small spin multiplets, i.e. those with $2j < r$ of the ambient space, while the multiplicities of the larger spin multiplets of the ambient space remain intact but become associated to a reduced spin: $j_\mathrm{ind}=j_\mathrm{amb}-\frac{r}{2}$. We will refer to this phenomenon as the {\it {Lefschetz cut}}.  This expression \eqref{Lefschetz-act-spin-funct-spin} also makes explicit that the only property of the vector bundle $\cale$ that plays a role in $\cali_\mathrm{ind} (\mm)$ is its rank, i.e. $r$. As we will see now the complete refined index $\cali(\mm)$ depends on various other properties of $\cale$.

\subsubsection{The complete index}
We already mentioned that due to its relation to the holomorphic Euler number the refined index can be computed by the Hirzebruch-Riemann-Roch theorem, see \eqref{rrh-index-thm}. Since $\calv$ is a vanishing locus of a section of a vector bundle $\cale$ over an ambient space $\calm$, the Todd class and Chern character appearing in that formula can be re-expressed in terms of classes associated to $\calm$ and $\cale$. Using the splitting principle and relations obtained via short exact sequences one finds
\begin{equation}
\cali (\mm \;;\;t ) = t^{r-d}\int_{\calm} \ttc \left(\calt\calm;t\right)\cdot  \frac{c_r (\mE)}{\ttc \left(\mE;t\right)}  \;, \label{vlv-holom-euler-numb-gen}
\end{equation}
where $c_r (\mE)$ is the top Chern class of $\mE$. 

\subsection{The special spaces of interest}\label{LSspec}
The ambient geometry associated to the quivers we study has a universal factor and it will be useful to specialize some of the general formulae of the previous subsection to this special case, as we'll see in the next subsection some physical features of interest are a direct consequence of this product structure. We thus specialize to the case where
\begin{equation}
\calm=\mCP^{a} \times \calk\qquad \cale= \oplus_k \left( H^\ast \otimes \mE_k \otimes \mbbc^{\x_k} \right)\label{specialcase}
\end{equation}
Here we assume that $\calk$ is a compact K\"ahler manifold, whose cohomology is diagonal in the sense of \eqref{cohas}. $H^*$ is the dual of the hyperplane bundle of the $\mCP^{a}$ and the factor $\mathbb{C}^{\xi_k}$ denotes the trivial vector bundle of rank $\xi_k$, while $\mE_k$ is a vector bundle over $\calk$ that is left unspecified for the moment. We introduce the additional shorthands:
\begin{equation}
\dim_\mathbb{C}\calk=d_0\qquad \mathrm{rank}_\mathbb{C}\, \mE_k=r_k
\end{equation}
It follows that $d=d_0+a$ and $r=\sum_k r_k\,\xi_k$. We will furthermore assume that
\begin{equation}
r\geq d_0
\end{equation}

\subsubsection{The ambient and induced index}
Both the ambient and induced index are fully determined in terms of the multiplicities $m_i$ of SU(2) irreps, see (\ref{kahler-spin-funct-spin}, \ref{Lefschetz-act-spin-funct-spin}). Using the decomposition of tensor products of SU(2) representations, i.e. the standard rules of composition of angular momentum, and the fact that the cohomology of $\mathbb{CP}^a$ forms a single spin $j=\frac{a}{2}$ multiplet one finds the following expression for the ambient multiplicities
\begin{align}
& m_{i}(\mathbb{CP}^a\times {\calk}) \qquad\  0\leq i\leq a+d_0 \label{mform}\\
&=\begin{cases} \sum_{k=\max(0,i-a)}^{\lfloor \frac{i-(a-d_0)}{2} \rfloor}m_{a+2k-i}({\calk})&\mbox{when }d_0\leq a\\
\Theta(2a-i)\sum_{k=\max(0,i-a)}^{\lfloor \frac{i}{2} \rfloor}m_{a+2k-i}({\calk})+\sum_{k=\max(0,\lceil \frac{i+a-d_0}{2}\rceil)}^{\min(a,\lfloor \frac{i-1}{2}\rfloor)}m_{a+i-2k} ({\calk})&\mbox{when }d_0> a
\end{cases}\nonumber
\end{align}
where $\Theta(x)$ is zero when $x<0$ and $1$ when $x\geq 0$. In general the induced multiplicities are
\begin{equation}
m_i(\mm)=m_{i+r}(\mathbb{CP}^a\times {\calk})\qquad 0\leq i \leq a+d_0-r\label{indind}
\end{equation}
but in a special range of parameters there is however the following identity:
\begin{equation}
m_{i+r}(\mathbb{CP}^a\times {\calk})=m_{i}(\mathbb{CP}^{a-r}\times {\calk})\quad\mbox{when}\quad r \leq a-d_0+1\,,\ 0\leq i \leq a+d_0-r
\end{equation}
This can be verified directly\footnote{In this verification it is useful to consider the cases $r \leq a-d_0$ and  $r = a-d_0+1$ separately.} from \eqref{mform}. Together with \eqref{indind} this implies that
\begin{equation}
\cali_\mathrm{ind}(\mm;t)=\chi_{a-r}(t)\cali(\calk;t)\quad\mbox{when}\quad r \leq a-d_0+1\label{indfact}
\end{equation}
Intuitively, what happens is that when $a$ is large enough the Lefschetz cut only slices through the cohomology of $\mCP^{a}$ leaving that of $\calk$ intact.

\subsubsection{The complete index}

We can also rewrite the general result (\ref{vlv-holom-euler-numb-gen}) for the class of geometries defined by \eqref{specialcase} as
 \begin{eqnarray}
\!\!\!\!\!\!\!\!\! \cali (\mm \;;\;t ) &=& t^{r-d}\!\!\int_{\calk} \int_{\mCP^{a}}\!\! \left[ \frac{h\, \left( 1- t\,e^{-h} \right) }{1-e^{-h}} \right]^{a+1}\!\!\!\!\!\!\!\! \ttc \left(\calt\calk;t\right) \prod_{k} \left( \frac{c_{r_k}(H^*\otimes \mE_k)}{\ttc \left(H^*\otimes\mE_k;t\right)}\right)^{\!\!\x_k}\label{indspec}
 \end{eqnarray}
Here $h$ is the first Chern class of $H^*$, which equals (minus) the K\"ahler class of $\mathbb{CP}^a$, in obtaining this formula the splitting principle is again the main tool.

\subsection{A criterion for the existence of intrinsic cohomology}\label{exist}
In this subsection we'll show that there is a simple requirement for ${\mathscr B}(\mm)$, the intrinsic contribution to the refined index as in \eqref{refindsplit}, to be non-vanishing, in the case the ambient manifold and defining vector bundle take the special form \eqref{specialcase} of the previous subsection. While this is presented as a purely mathematical observation here, we will discuss in the later sections  the important physical consequence of this criterion that, for the quivers we will consider, pure-Higgs states only exist in the scaling regime.

The precise statement is that under the conditions of the previous subsections:
\begin{quote}
	{\bf Existence criterion} : If ${\mathscr B}(\mm;t)\neq 0$ then $r-d_0<a+1<d_0+r$\,.
\end{quote}
The lower bound is rather trivial, since $\calv$ is non-empty only if $a+d_0-r\geq 0$. The upper bound can also be read as saying that ${\mathscr B}(\mm;t)=0$ when $a+1\geq r+d_0$, which in turn is equivalent to $\cali(\mm;t)=\cali_\mathrm{ind}(\mm;t)$ when $a+1\geq r+d_0$. It is this last equality that we now prove. As pointed out above, when $a+1\geq r+d_0$ the induced index takes the product form \eqref{indfact}, it remains thus to show that also $\cali(\mm;t)$ takes this form. We start by further working out \eqref{indspec}:
\begin{equation}
\cali (\mm ;t ) = t^{r-d_0-a} \,\int_{\calk} \oint \frac{dy}{1-(1-t^2) \, y} \;\left( \frac{ 1+ t^2\,y}{y} \right)^{a+1}\!\!\!\!\!\!\! \ttc(\calt\calk;t) \prod_k\prod_{i=1}^{r_k} \left( \frac{y + z_{ik}}{1 + t^2 \, y + z_{ik}  } \right)^{\xi_k} \;, \label{proto-ref-ind-gen}
\end{equation}
Here we used that the integration over $\mathbb{CP}^a$ in \eqref{proto-ref-ind-gen} simply selects the coefficient of $h^a$ and made the change of variables $e^{-h}=1-(1-t)y$. In addition we introduced the forms $z_{ik}  = \frac{-1}{1-t^2} \, \left(1 - e^{-x_{ik}} \right)$
with $x_{ik}$, $i=1,\ldots, r_k$, the Chern roots of $\mE_k$. 


To show factorization when $a+1\geq r+d_0$ we first make the key observation that
\begin{equation}
\!\!\!\!\!  \oint \frac{t^{r-a}\, dy}{1-(1-t^2) \, y}\left( \frac{ 1+ t^2\,y}{y} \right)^{a+1}\!\!\! \prod_{i,k} \left( \frac{y + z_{ik}}{1 + t^2 \, y + z_{ik}  } \right)^{\xi_k}  = \,\chi_{a-r}(t)+\calo(z^{a-r+2})\quad \mbox{when }a\geq r\label{cruxeq}
\end{equation}
where we consider $\calo(z_{ik})=\calo(z)$. This formula is obtained in two steps. First one can verify that
\begin{equation}
\chi_n(t)=\oint \frac{t^{n}\, dy}{1-(1-t^2) \, y}\left( \frac{ 1+ t^2\,y}{y} \right)^{n+1}\label{intchar}
\end{equation}
Secondly we remark that it is sufficient to prove \eqref{cruxeq} for $z_{ik}=z$, which can be done by a direct computation using \eqref{intchar}:
\begin{align*}
&\chi_{a-r}(t)-\oint \frac{t^{r-a}\, dy}{1-(1-t^2) \, y}\left( \frac{ 1+ t^2\,y}{y} \right)^{a+1}\left( \frac{y + z}{1 + t^2 \, y + z  } \right)^{r}\\
 &= \oint \frac{t^{r-a}\, dy}{1-(1-t^2) \, y}\frac{(1+ t^2\,y)^{a-r+1}}{y^{a+1}(1+t^2y+z)^r}\left((1+t^2y+z)^ry^r-(y + z)^r(1+t^2y)^r\right)\\
 &=-\left(rz+\calo(z^2)\right)\oint \frac{t^{r-a}\, dy}{y^{a-r+2}}\frac{(1+t^2 y)^a}{((1+t^2y+z)^r}\\
 &=-\left(rz+\calo(z^2)\right)\frac{t^{a-r+2}z^{a-r+1}}{(1+z)^{a+1}}\begin{pmatrix}
 a\\ r-1
 \end{pmatrix}=\calo(z^{a-r+2})
\end{align*}
The estimate \eqref{cruxeq} is sufficient for our purposes since upon insertion into \eqref{proto-ref-ind-gen} one sees that the $\calo(z^{a-r+2})$ part vanishes upon integration over $\calk$ when $a-r+2>d_0$. Using \eqref{proto-ref-ind-gen} for the remaining part we then get
\begin{equation}
\cali (\mm\;;\;t) = \chi_{a-r}(t) \; \cali (\calk \;;\; t)\qquad\mbox{ when }\ a+1\geq d_0+r \label{eq2}
\end{equation}
This equality between the total refined index \eqref{eq2} and its induced part \eqref{indfact} then concludes the derivation of the existence criterion above.

\subsection{A factorization formula}\label{facsec}
\begin{figure}
	\centering
	\includegraphics[scale=0.5]{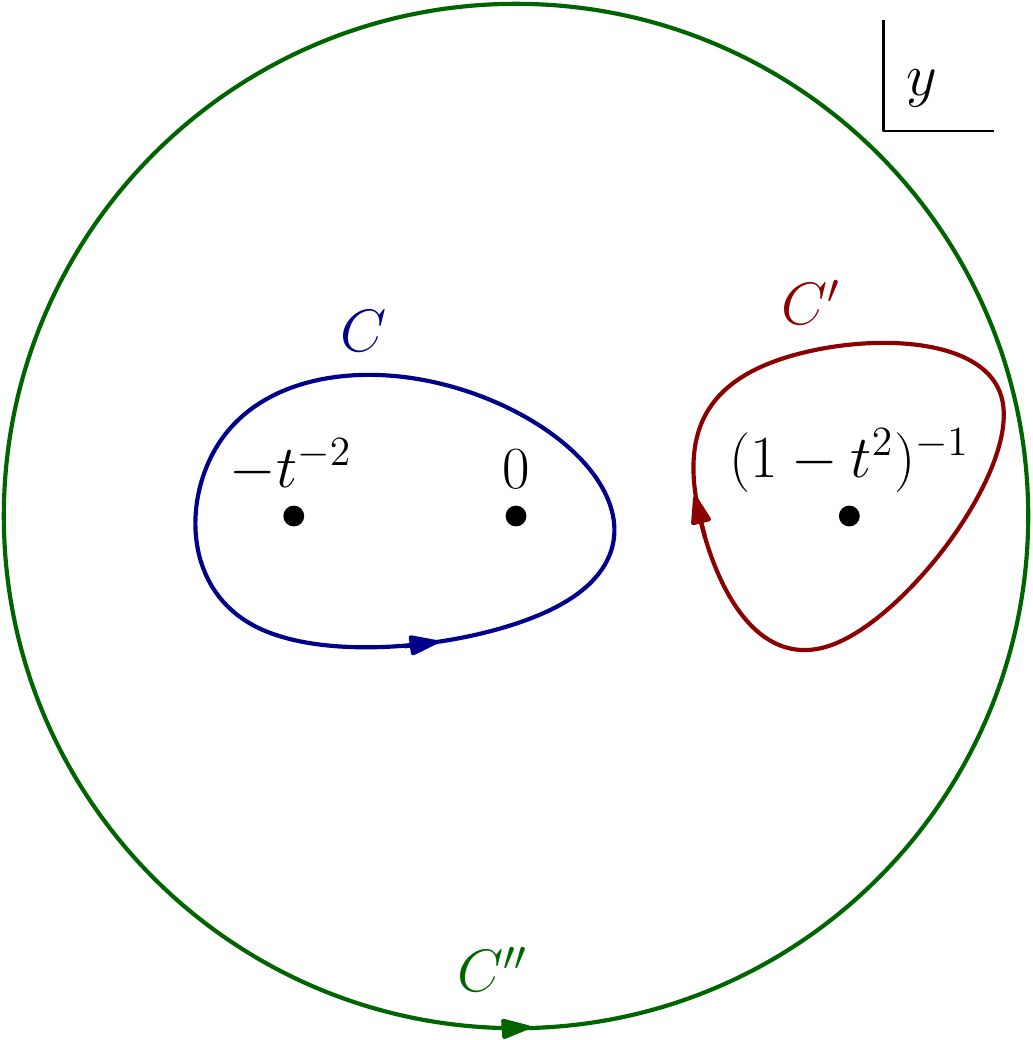}\caption{The three contours of relevance in the computation of $\Delta \cali$.\label{contfig}}
\end{figure}

In the previous subsection we saw that for appropriate values of parameters the induced index and total index of the manifolds we consider are equal and furthermore factorize, as in \eqref{eq2}. We now show that a similar relation exists for all values of the parameters, as long as we consider not the indices themselves but some 'difference' related to them. In the next section we will see how this difference is related to wall-crossing between two chambers of the quiver. Let us first abstractly define the two 'differences' we are interested in
\begin{equation}
\Delta \cali(\mm;t)= t^{r-d_0-a} \,\int_{\calk} \oint_C \frac{dy}{1-(1-t^2) \, y} \;\left( \frac{ 1+ t^2\,y}{y} \right)^{a+1}\!\!\!\!\!\!\! \ttc(\calt\calk;t) \prod_k\prod_{i=1}^{r_k} \left( \frac{y + z_{ik}}{1 + t^2 \, y + z_{ik}  } \right)^{\xi_k} \;, \label{deltotind}
\end{equation}
With the contour $C$ a closed curve around the points $y=0$ and $y=-\frac{1}{t^2}$ that excludes the point $y=(1-t^2)^{-1}$, see figure \ref{contfig}. Note that by this definition $\Delta \cali(\mm,t)$ includes $\cali(\mm,t)$ from the contribution around $y=0$ as in \eqref{proto-ref-ind-gen} but also has an extra contribution from the pole at $y=-t^2$, since we should consider the expressions containing $z$'s as formal power series in these variables. i.e. $(1+t^2y+z)^{-1}=\sum_{n} z^n (1+t^2y)^{-n-1}$. 

For the induced index we write the difference more explicitly. Because $\mm$ is the vanishing locus of a section of a rank $r$ vector bundle over $\mathbb{CP}^{a}\times \calm_0$ let us write $\cali_\mathrm{ind}(\mm)=\cali_\mathrm{ind}(a,r)$. Using this notation we define
\begin{equation}
\Delta \cali_\mathrm{ind}(\mm;t)=\cali_\mathrm{ind}(a,r;t)-\cali_\mathrm{ind}(r-1,a+1;t) \label{dinddef}
\end{equation}
From the definition (\ref{deltotind}, \ref{dinddef}) it follows that
\begin{quote}
	{\bf Factorization} : $ \Delta \cali(\mm,t)=\Delta \cali_\mathrm{ind}(\mm;t)=\chi_{a-r}(t)\cali(\calm_0;t)$ \label{factform}
\end{quote} 
Note that this equality is valid for $r>a$ as well, in which case the actual SU(2) characters are obtained via the analytic continuation
\begin{equation}
\chi_{-n}(t)=-\chi_{n-2}(t) \quad \mbox{for } n<0\label{negspin}
\end{equation}. Before we argue why the factorization formula above is correct, let us point out that this also implies that
\begin{equation}
\Delta {\mathscr B}=\Delta \cali-\Delta\cali_{\mathrm{ind}}=0  \label{Hggs-no=jump}
\end{equation}

To derive the factorization of $\Delta \cali_\mathrm{ind}$ it is useful to first re-express the indices in terms of Betti-numbers. Via \eqref{poicare-polynom-spin-coeff-1} it follows that
\begin{equation}
 \cali (\mathbb{CP}^a\times\calm_0) = \sum_{\ell=0}^{2\,d_0} b^{(\ell)} (\calk) \, \chi_{a-d_0+\ell} (t)  \;. \label{amb-manif-spin-funct} 
\end{equation}
the SU(2) characters with negative index should again be interpreted via \eqref{negspin}. Via the Lefschetz cut \eqref{Lefschetz-act-spin-funct-spin} this leads to
\begin{equation}
\cali_\mathrm{ind}(a,r) = \sum_{k=0}^{d_0+a-r} b^{(k)} (\calk) \, \chi_{d_0+a-r-k} (t) + \sum_{k=d_0+a+r+1}^{\infty} b^{(k)} (\calk) \; \chi_{d_0+a+r-k} (t) \;. \label{vac-spin-funct-case-generic}
\end{equation}
Factorization then follows from the definition \eqref{dinddef} after a few additional steps of algebra.

We can compute  $\Delta \cali$ via a contour argument. As illustrated in figure \ref{contfig} we can replace the integral around $C$ with the sum of the integrals along $C'$ and $C''$. The first integral simply reduces to (minus) the residue of the integrand at $y=(1-t^2)^{-1}$, while we can compute the second integral by sending the radius of the circle $C''$ to infinity. One finds
\begin{eqnarray}
\oint_{C'} \frac{dy}{1-(1-t^2) \, y} \;\left( \frac{ 1+ t^2\,y}{y} \right)^{a+1}\prod_k\prod_{i=1}^{r_k} \left( \frac{y + z_{ik}}{1 + t^2 \, y + z_{ik}  } \right)^{\xi_k}&=&\frac{1}{1-t^2}\\
\oint_{C''} \frac{dy}{1-(1-t^2) \, y} \;\left( \frac{ 1+ t^2\,y}{y} \right)^{a+1}\prod_k\prod_{i=1}^{r_k} \left( \frac{y + z_{ik}}{1 + t^2 \, y + z_{ik}  } \right)^{\xi_k}&=&\frac{t^{2(a-r+1)}}{t^2-1}
\end{eqnarray}
which combined with \eqref{deltotind} leads directly to the factorization formula above. 
 
\section{The $2+n$ node quivers} \label{sect-3nodes-hggs-branch}
 $\caln=4$ quiver quantum mechanics \cite{Denef:2002ru}, is described by a quiver, which is a collection of nodes -- labeled by an index $\alpha$ say -- with arrows connecting them. The number of arrows between the nodes labeled by $\alpha$ and $\beta$ is denoted as $\gamma_{\alpha\beta}$, and this defines an anti-symmetric matrix by letting the sign indicate the direction of the arrows. There is a unitary gauge group associated to each node and each arrow comes with a complex scalar field transforming in the bi-fundamental representation of the gauge groups of the nodes the arrows connect. In addition there are three adjoint valued scalar fields for each node, apart from a number of gauge fields, fermions and non-dynamical fields that render the theory invariant under supersymmetry and gauge transformations. In this paper we'll restrict attention to a special class of quivers, that we'll call $\mathit{2+n}$ {\it node quivers}.
 
 The $2+n$ node quivers have by definition 2 abelian nodes and $n$ non-abelian nodes. We'll refer to the two abelian nodes as $0$ and $\bar 0$  respectively, while we'll label the different non-abelian nodes with an index $\mu=1,\ldots, n$. The unitary gauge group at each of the non-abelian nodes is allowed to have an arbitrary rank, $N_\mu$ respectively\footnote{We should point out that although the gauge group at the 'non-abelian' nodes is typically indeed non-abelian, the special case where some of the $N_\mu=1$ is included, in which case some of the 'non-abelian' nodes actually carry an abelian gauge group.}. We'll take as part of the definition of a $2+n$ node quiver a strong restriction on the number and type of arrows, the only non-zero entries of the matrix describing them are (up to anti-symmetry)
\begin{equation}
\gamma_{0\bar 0}=\kappa_0\qquad \gamma_{\bar 0 \mu}=\gamma_{\mu 0}=\kappa_\mu
\end{equation}
Here all $\kappa$'s are assumed to be strictly positive integers, which has the important implication that every $2+n$ node quiver has a number of closed loops. The diagram of a $2+n$ node quiver is sketched in figure \ref{2nquiv}. 

This class of $2+n$ node quivers is motivated by the D4D0 system of type II string theory Calabi-Yau compactifications. If one considers multi-center configurations made of a (fluxed) D6 brane of charge $\Gamma_{\mathrm{D}6}$, a (fluxed) anti-D6 brane of charge $\Gamma_{\overline{\mathrm{D}6}}$ and a number of (anti) D0 branes \cite{Denef:2007vg}, namely $N_\m$ multiple D0 branes of charge $\Gamma_\mu=\mu \Gamma_{\mathrm{D0}}$, then this has a description in terms of a $2+n$ node quiver with
\begin{equation}
\kappa_0=\langle \Gamma_{\mathrm{D}6}, \Gamma_{\overline{\mathrm{D}6}}\rangle\qquad \kappa_\mu=\mu\, \langle \Gamma_{\mathrm{D0}}, \Gamma_{\mathrm{D}6}\rangle
\end{equation}

Just as any quiver quantum mechanics, the $2+n$ node quivers have a Coulomb and a Higgs branch.

The Coulomb branch of $\caln=4$ quiver quantum mechanics can be given the interpretation of a number of dyonic points located in $\mathbb{R}^3$. In the case of a $2+n$ node quiver, we have points $\mathbf{x}_0$ and $\mathbf{x}_{\bar 0}$ for each the abelian nodes and points $\mathbf{x}_{\mu a}$, $a=1,\ldots, N_\m$ for each non-abelian node. Denoting the distance between the various points respectively as $r_{0\,\bar 0}$, $r_{0\,\mu i}$ and $r_{\bar 0\,\mu i}$, the Coulomb branch can be identified with the solutions to the Denef equations
\begin{eqnarray}
\frac{\kappa_0}{r_{0\,\bar 0}}-\sum_{\mu,a}\frac{\kappa_\m}{r_{0\,\mu a}}&=&-2\theta_0\nonumber\\
-\frac{\kappa_0}{r_{0\,\bar 0}}+\sum_{\mu,a}\frac{\kappa_\m}{r_{\bar 0\,\mu a}}&=&-2\theta_{\bar 0}\label{Cbranch}\\
\frac{\kappa_\m}{r_{0\,\mu a}}-\frac{\kappa_\m}{r_{\bar{0}\,\mu a}}&=&-2\theta_\m\nonumber
\end{eqnarray}
There is a natural angular momentum associated to each solution, physically originating from the electromagnetic field of the dyons. Quantization of this angular momentum leads to the quantum states on the Coulomb branch \cite{deBoer:2008zn}. 

The Higgs branch of $\caln=4$ quiver quantum mechanics is  a quotient by the gauge group of the set of solutions to the D- and F-term equations. For the $2+n$ node quivers we have the complex scalars $\varphi_{0\bar0}{}^{k}$, $k=1,\ldots ,\kappa_0$ and $\varphi_{0\mu}{}^{ka}$, $\varphi_{\bar 0\mu}{}^{k}{}_a$, with $k=1,\ldots \kappa_\m$ and the upper (lower) index $a=1\ldots N_\m$ an (anti-)fundamental $U(N_\m)$ index. The equations defining the Higgs branch then read
\begin{eqnarray}
\sum_{k=1}^{\k_\m} (\varphi_{0\m}{}^{ka}\bar \varphi_{0\m}{}^k{}_b-  \bar\varphi_{\bar 0\m}{}^{ka}\varphi_{\bar 0\m}{}^k{}_b)&=&-\theta_\m \delta^a_b \nonumber\\
\sum_{k=1}^{\k_0}\bar \varphi_{0\bar 0}{}^k \varphi_{0\bar 0}{}^{k}-\sum_{\m=1}^n\sum_{k=1}^{\kappa_\m}\sum_{a=1}^{N_\m}\bar \varphi_{0\m}{}^k{}_a\varphi_{0\m}{}^{ka}&=&-\theta_0 \nonumber\\
-\sum_{k=1}^{\k_0}\bar \varphi_{0\bar 0}{}^k \varphi_{0\bar 0}{}^{k}+\sum_{\m=1}^n\sum_{k=1}^{\kappa_\m}\sum_{a=1}^{N_\m} \varphi_{\bar 0\m}{}^k{}_a\bar \varphi_{\bar 0\m}{}^{ka}&=&-\theta_{\bar 0}\label{HiggsBranch}\\
\sum_{k_1=1}^{\kappa_0}\sum_{k_2=1}^{\kappa_\mu}\, C_{\m k k_1 k_2}\varphi_{0\bar 0}{}^{k_1}\varphi_{\bar 0 \m}{}^{k_2}{}_a&=&0\nonumber\\
\sum_{k_1=1}^{\kappa_0}\sum_{k_2=1}^{\kappa_\mu} C_{\m k_2 k_1 k}\,\varphi_{0 \m}{}^{k_2 a}\varphi_{0\bar 0}{}^{k_1}&=&0\nonumber\\
\sum_{\mu=1}^n\sum_{k_1=1}^{\kappa_\m}\sum_{k_2=1}^{\kappa_\mu}\sum_{a=1}^{N_1}\, C_{\m k_1 k k_2}\varphi_{0 \m}{}^{k_1 a}\varphi_{\bar 0 \mu}{}^{k_2}{}_a&=&0\nonumber
\end{eqnarray}
The first three equations are the D-terms, while the last three equations are F-terms that originate from a gauge invariant holomorphic superpotential that without loss of generality can be chosen to be cubic \cite{Denef:2007vg}. The supersymmetric quantum states on the Higgs branch are described by its Dolbeault cohomology.

Note that both the Coulomb and Higgs branch depend on the choice of Fayet-Iliopoulos (FI) parameters $\theta_0$, $\theta_{\bar 0}$ and $\theta_\m$. We will refrain from exploring the full range of possible FI parameters. Instead we will restrict attention to special values that describe two chambers, which is enough for our purpose to check the stability of the pure-Higgs states. A generic choice of FI parameters would allow any $\theta_0$, $\theta_{\bar 0}$ and $\theta_\mu$ subject to the constraint $\theta_0+\theta_{\bar 0}+\sum_{\m=1}^{n} N_\m \theta_\mu=0$, in this paper we will however look at the special case
\begin{equation}
\theta_0=\theta -\frac{N\epsilon}{2}\,\qquad \theta_{\bar 0}=-\theta -\frac{N\epsilon}{2}\qquad \theta_\mu=\epsilon\qquad\quad N=\sum_{\mu=1}^{n}N_\m\label{FI}
\end{equation}
The situation $\epsilon=0$ is most symmetric, and corresponds to a point of {\it threshold stability} \cite{deBoer:2008fk} from the point of view of the D6, anti-D6, D0 system and the choice $\theta>0$ then connects to the regime where an AdS$_2\times$S$^3$ decoupling limit exists \cite{deBoer:2008fk}. The choice $\epsilon=0$ is however intrinsic from the geometric perspective, that is why we will choose $\epsilon\neq 0$ but consider it to be infinitesimally small for all purposes. Choosing $\epsilon$ positive or negative will lead to identical results -- this is what distinguishes a wall of threshold stability from one of marginal stability -- but for clarity we will restrict attention to $\epsilon>0$. The parameter $\theta$ is allowed to take any\footnote{The quiver description is only of physical relevance when $\theta$ is small enough, but on its own it is well defined also for larger $\theta$.} real value. As we will see below there are two chambers, separated by a wall of marginal stability at $\theta=\frac{N\epsilon}{2}$, it will be useful to label them:
\begin{quote}
	{\bf negative chamber} :\quad $\theta <  \frac{N\epsilon}{2}$\qquad\qquad 	{\bf positive chamber} :\quad $\theta >  \frac{N\epsilon}{2}$
\end{quote}
Strictly speaking $\theta$ does not have to be negative in the negative chamber, but since we assume $\epsilon$ to be infinitesimal $\theta$ is  'effectively' negative in that chamber.

We now have the setup ready to start our investigation of the states of the $2+n$ node quivers and some of their properties. Although we have results for arbitrary $n$, notation and combinatorics might cloud some of the main points in this general case. For reasons of presentation we will therefore first discuss in detail the case $n=1$, while keeping the discussion for arbitrary $n$ shorter and focused on results, since the main intuition from the $n=1$ case carries over quite directly modulo some extra combinatorics.

\subsection{$n=1$}
\begin{figure}[tbp]
	\centering
	\includegraphics[scale=1]{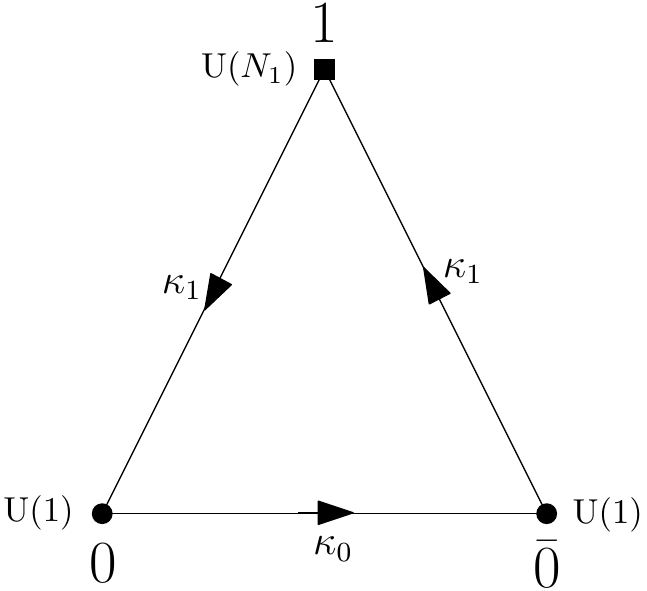}
	\caption{The quiver diagram of a $2+1$ node quiver.}\label{21quiv}
\end{figure}

The $2+1$ node quiver is a special 3 node quiver, as illustrated in the diagram of figure \ref{21quiv}, and is completely characterized by the three integers $\kappa_0$, $\kappa_1$ and $N_1$. Both the equations defining the Coulomb and Higgs branch, \eqref{Cbranch} and \eqref{HiggsBranch}, simplify because we can put $\mu=1$. We'll discuss each in turn. 
\subsubsection{Coulomb branch states} \label{section-n1-coulmb}
Our aim will be to count the states on the Coulomb branch (CB), keeping track of their total angular momentum. Letting $m^\mathrm{tot}$ denote the $J_3^{\mathrm{total}}$ eigenvalue of a state, we then formally define
\begin{equation}
\cali_\mathrm{C}(\kappa_0, \kappa_1, N_1;t)=\sum_{\mbox{\scriptsize{states on CB}}} t^{2m^\mathrm{tot}}\label{coulind}
\end{equation}
An enumeration of states on the Coulomb branch under consideration was already performed in \cite{deBoer:2009un} and we refer to that work for a details, we will only recall the main steps and compute the index \eqref{coulind}.

As we mentioned before quantization of the Coulomb branch is essentially a quantization of the angular momenta of dyonic particles. In the case at hand we haven an angular momentum for each of the $N_1$ particles associated to the non-abelian node, let's call them $\vec{J}^a$, and an angular momentum associated to the two abelian nodes, let's call it $\vec{J}^0$. Instead of keeping track of $\vec{J}^0$ it is more convenient to work with $\vec{J}^a$ and $\vec{J}^\mathrm{tot}=\vec{J}^a+\vec{J}^0$. We will denote the quantum numbers of $J_3^a$ and $J_3^{\mathrm{total}}$ as $m^a$ and $m^\mathrm{tot}$ respectively. The constraints on these quantum numbers originate in the constraints on the positions of the particles, described by the Denef equations \eqref{Cbranch}. They take the form\footnote{In \cite{deBoer:2009un} only the negative chamber is considered, but the constraints of the positive chamber become identical to those of the negative chamber upon the replacement $m^a\rightarrow \kappa_1-m^a$ and $\kappa_0\rightarrow 2 \kappa_1 N_1-\kappa_0$.} \cite{deBoer:2009un}:
\begin{align}
0 \leq m^1 < m^2 < \ldots < m^{N_1} < \kappa_1  \;, \qquad -j^\mathrm{tot}\leq m^\mathrm{tot}\leq j^\mathrm{tot}   \label{dyonic-ang-mom-descipt}
\end{align}
where
\begin{equation}
j^\mathrm{tot}= \frac{\k_\theta-N_1-1}{2} - \sum_{a=1}^{N_1} m^a\quad\mbox{and}\quad \kappa_\theta=\begin{cases}
\kappa_0 &\quad \theta \in \mbox{negative chamber}\\
2\kappa_1 N_1-\kappa_0 &\quad \theta \in \mbox{positive chamber}
\end{cases}\label{defjk}
\end{equation} To count the number of states it is useful to make the redefinition $\hat m^a=m^a-a+1$, then
\begin{equation}
0\leq \hat m^1\leq \hat m^2\leq \ldots \leq \hat m^{N_1}\leq \kappa_1-N_1\qquad j^\mathrm{tot}= \frac{\k_\theta-N_1^2-1}{2} - \sum_{a=1}^{N_1} \hat m^a\label{jtot2}
\end{equation} 
Note that while we trade the $m^a$ for $\hat m^a$, we keep $m^\mathrm{tot}$ and the second condition in \eqref{dyonic-ang-mom-descipt} intact, this in particular implies $j^\mathrm{tot}$ needs to be positive.
Let us now fix for a moment $M=\sum_{a=1}^{N_1} \hat m^a$, then we see that a choice of $\hat m^a$ corresponds to a partition of $M$ into $N_1$ terms, each of which is smaller than $\kappa_1-N_1$. The number of such partitions, let us call it $P_M^{(\kappa_1,N_1)}$, is a well known combinatorial object whose generating function is the $q$-binomial:
\begin{equation}
\sum_{M=1}^\infty P_M^{(\kappa_1,N_1)} q^M=\begin{bmatrix}\kappa_1\\N_1\end{bmatrix}_q=\frac{\prod_{i=1}^{\kappa_1}(1-q^i)}{\prod_{i=1}^{N_1}(1-q^i)\prod_{i=1}^{\kappa_1-N_1}(1-q^i)}\label{qbin}
\end{equation}
Also for fixed $M$ we see that $m^\mathrm{tot}$ simply spans a spin $j^\mathrm{tot}$ multiplet. So we find
\begin{equation}
\left.\cali_\mathrm{C}(\kappa_0,\kappa_1,N_1;t)\right|_{M\mbox{ \scriptsize{fixed}}}=P_M^{(\kappa_1,N_1)} \chi_{2j^\mathrm{tot}}(t)
\end{equation}
To get the full answer we simply need to sum this over all possible values of $M$. Since $j^\mathrm{tot}$ needs to be non-negative for there to be solutions we see that $2M\leq \k_\theta-N_1^2-1$ and thus
\begin{equation}
\cali_\mathrm{C}(\kappa_0,\kappa_1,N_1;t)=\sum_{M=0}^{\lfloor\frac{\k_\theta-N_1^2-1}{2}\rfloor}P_M^{(\kappa_1,N_1)} \chi_{\k_\theta-N_1^2-1-2M}(t)\label{Cgen1}
\end{equation}
Although correct, the above expression might not be the most practically useful. We observe that has the form of a discrete convolution, which can be transformed into a standard product by introducing a generating function conjugate to $\kappa_\theta$, i,e. we define 
\begin{equation}
\mathbb{I}_\mathrm{C}(\kappa_1,N_1; t,z)=\sum_{\kappa_\theta=0}^\infty \cali_\mathrm{C}(\kappa_0,\kappa_1,N_1;t) z^{\kappa_\q}  \label{coulomb-refined-indx-gen-funct-def}
\end{equation}
Using \eqref{qbin} and $\sum_{n=0}^\infty \chi_n(t)z^n=(1-tz)^{-1}(1-t^{-1}z)^{-1}$ one finds
\begin{equation}
\mathbb{I}_\mathrm{C}(\kappa_1,N_1; t,z)=\frac{z^{N_1^2+1}}{(1-tz)(1-t^{-1}z)}\begin{bmatrix}\kappa_1\\N_1\end{bmatrix}_{z^2} \label{coulomb-refined-indx-gen-funct-exprn}
\end{equation}
This can be transformed back to provide an alternative form for \eqref{Cgen1}:
\begin{equation}
\cali_\mathrm{C}(\kappa_0,\kappa_1,N_1;t)=\oint dz\frac{z^{N_1^2-\kappa_\theta}}{(1-tz)(1-t^{-1}z)}\begin{bmatrix}\kappa_1\\N_1\end{bmatrix}_{z^2}\label{Cgen2}
\end{equation}
Note that this result depends on $\theta$ and takes different values in the two chambers, which is a manifestation of wall-crossing, it is consistent with the Coulomb branch formula \cite{Manschot:2014fua} as can be verified by the Mathematica package \texttt{CoulombHiggs.m} \cite{CHpioline}. We will investigate the precise jump further below when we discuss the Higgs branch. We finish our discussion of the Coulomb branch with two related observations.
 
Since the integrand of \eqref{Cgen2} is holomorphic at the origin whenever $N_1^2-\k_\theta\geq 0$ the contour integral vanishes in that case, which means the Coulomb branch does not support any supersymmetric ground states. Using the definition \eqref{defjk} of $\kappa_\theta$ in the two chambers we can summarize this observation as
\begin{quote}
	{\bf Coulomb existence criterion}: The Coulomb branch is empty\\
	 {}\qquad \hspace{10cm}a) in the negative chamber when $\kappa_0 \leq N_1^2$,\\
	 \qquad b) in the positive chamber when $\kappa_0 \geq N_1(2\kappa_1-N_1)$.
\end{quote}
The parameter regime where the Coulomb branch is non-empty in both chambers is closely related to another phenomenon, that of the existence of a {\it scaling point} \cite{Bena:2006kb, Denef:2007vg, Bena:2007qc}, where the centers can approach each other arbitrarily close in solution space. This classical notion gets corrections due to the quantum spin of the centers, which is taken into account for the states that we enumerated above. Since the classical scaling solutions have zero angular momentum a necessary condition for the presence of a scaling solution among the states counted above would be for the naive\footnote{Of course being the size of an angular momentum $j^\mathrm{tot}$ is always positive, which is an extra constraint on the $m^a$.} minimal value of $j^\mathrm{tot}$ in \eqref{jtot2} to be smaller or equal to 0.  One checks that this is equivalent to the condition $N_1^2<\kappa_0\leq N_1(2\kappa_1-N_1)+1$ in the negative chamber and $N_1^2-1\leq \kappa_0< N_1(2\kappa_1-N_1)$ in the positive chamber. One sees that the overlap of these two charge regions coincides with the condition for the Coulomb branch to exist/be non-empty in both chambers. This is a slightly different, but probably better characterization of the scaling regime with quantum corrections included, since as follows from e.g. the Denef equations in the scaling limit, a crucial property that distinguishes scaling solutions from more ordinary multicenter solutions is that their existence is independent of the value of the FI parameters and depends on the 'charge' parameters only. This thus motivates us to define the (quantum corrected) scaling regime as
\begin{quote}
{\bf Scaling regime}: The Coulomb branch is non-empty in both chambers iff
\begin{equation}
N_1^2<\kappa_0<N_1(2\kappa_1-N_1)\label{scalreg}
\end{equation}
\end{quote}
Although this regime can be defined from knowledge of the Coulomb branch only, we will show below that it is also exactly the regime where pure-Higgs states are present.

In the case where the quiver is fully abelian, i.e. $N_1=1$, there is the previously studied quantum corrected scaling condition $a\leq b+c-2$ (and cyclic) \cite{deBoer:2012ma, Manschot:2012rx}, one easily verifies that \eqref{scalreg} reproduces this when $N_1=1$, $a=\kappa_0$ and $b=c=\kappa_1$.

\subsubsection{Higgs branch states} \label{sect-Higgs-n1}
Although the equations \eqref{HiggsBranch} defining the Higgs branch look somewhat daunting, there is a powerful argument \cite{Denef:2007vg} that greatly simplifies their analysis. The upshot is that if the numbers $C_{1 k_1 k_2 k_3}$ are generic -- which we'll assume -- then every solution to the last three equations of \eqref{HiggsBranch} must have at least one of $\varphi_{00}$, $\varphi_{01}$ or $\varphi_{\bar 0 1}$ zero. What fields can vanish is in turn restricted by the first three equations and the choice of FI parameters. For this reason we need to split our analysis per chamber.

\paragraph{The negative chamber}
In this chamber we have $\theta_0<0$ and $\theta_1>0$ and one sees that the first three equations of \eqref{HiggsBranch} only allow a solution if we make the choice $\varphi_{01}{}^{ka}=0$. The equations then simplify to
\begin{eqnarray}
\sum_{k=1}^{\k_1} \bar\varphi_{\bar 01}{}^{ka}\varphi_{\bar 01}{}^k{}_b&=&\theta_1 \delta^a_b \label{HBneg1}\\
\sum_{k=1}^{\k_0}\bar \varphi_{0\bar 0}{}^k \varphi_{0\bar 0}{}^{k}&=&-\theta_0 \label{HBneg2} \\
\sum_{k_1=1}^{\kappa_0}\sum_{k_2=1}^{\kappa_1}\, C_{1 k k_1 k_2}\varphi_{0\bar 0}{}^{k_1}\varphi_{\bar 0 \m}{}^{k_2}{}_a&=&0\label{HBneg3}
\end{eqnarray}
Note that the first two equations are decoupled, and can be solved before considering the third. Geometrically this means that the first two equations define an ambient space $\calm$, which will be a direct product, while the third equation will select a subspace $\calv$ inside the ambient space. The solutions to \eqref{HBneg2}, upon a quotient by U(1) gauge symmetry, parameterize the projective space $\mathbb{CP}^{\kappa_0-1}$. Similarly, once quotiented by U($N_1$), the solutions to the set of equations \eqref{HBneg1} describe $\Gr(N_1,\kappa_1)$, the Grassmannian manifold parameterizing $N_1$-planes in $\mathbb{C}^{\kappa_1}$. In summary
\begin{equation}
\calm_-=\mathbb{CP}^{\kappa_0-1}\times \Gr(N_1,\kappa_1)\,.\label{negamb}
\end{equation}
Mathematically speaking the left hand side of the remaining set of equations \eqref{HBneg3} defines a section of a vector bundle $\cale_-$ over $\calm_-$. It is important to note that $\varphi_{0\bar 0}$ and $\varphi_{\bar0 1}$ are not precisely coordinates on $\calm_-$. Rather $\varphi_{0\bar 0}$ is itself a section of $H$, the hyperplane bundle  of $\mathbb{CP}^{\kappa_0-1}$, while $\varphi_{\bar0 1}$ is a section of $S$, the dual of the tautological bundle of $\Gr(N_1,\kappa_1)$.  Since the left hand side of \eqref{HBneg3} takes the sections $\varphi_{0\bar 0}$ and $\varphi_{\bar0 1}$ to a complex number for each choice of the free index $k=1,\ldots,\kappa_1$, we can identify the vector bundle that this left hand side is a section of, as
\begin{equation}
\cale_-=H^*\otimes S^*\otimes\mathbb{C}^{\kappa_1}\label{negbund}
\end{equation}
We thus conclude that in the negative chamber the Higgs branch can be identified with $\calv_-$, the vanishing locus of a section of the vector bundle $\cale_-$ over the ambient manifold $\calm_-$, which puts it into the geometric setting of section \ref{LSsec}.  In particular the Higgs branch in this chamber is an example of the special class of spaces discussed in section \ref{LSspec} with the geometric parameters of that section being
\begin{equation}
a=\kappa_0-1\qquad d_0=N_1(\kappa_1-N_1)\qquad r=\kappa_1 N_1\label{parmap1}
\end{equation}
Note that this automatically implies $r\geq d_0$. Since the relevant properties of $\calv_-$ are fully determined in terms of the integers $\kappa_0,\kappa_1$ and $N_1$, we will -- in this part of the paper -- thus often write $\cali(\calv_-)=\cali^-(\kappa_0,\kappa_1,N_1)$ etc.

 The mathematical identification of $\calv$ we just made implies the cohomology of $\calv$, i.e. the supersymmetric ground states on the Higgs branch, splits into an induced part and an intrinsic part, as discussed in section \ref{LSsec}. At the level of the refined index this split can be written as \eqref{refindsplit}. To compute $\cali_\mathrm{ind}$, see \eqref{Lefschetz-act-spin-funct-spin}, we only need knowledge of the cohomology of the ambient manifold and keep track of the Lefschetz cut. The cohomology of projective space and the Grassmanian are well known. On $\mathbb{CP}^{\k_0-1}$ there is a unique class at each even degree between $0$ and $\kappa-1$, while on $\Gr(N_1,\kappa_1)$ the cohomology at degree $2M$ is in one to one correspondence to partitions of $M$ into $N_1$ parts, each smaller than $\kappa_1-N$. We already encountered these partitions on the Coulomb branch, and this is no coincidence. Working out \eqref{Lefschetz-act-spin-funct-spin} in detail for \eqref{negamb} one finds
 \begin{equation}
 \cali_{\mathrm{ind}}^-(\kappa_0,\kappa_1,N_1;t)=\cali_{\mathrm{C}}^{-}(\kappa_0,\kappa_1,N_1;t)\label{indiscoul}
 \end{equation}
 with the right hand side given in \eqref{Cgen1}, where one should remember that in the negative chamber $\kappa_\theta=\kappa_0$. This identification of the induced cohomology on the Higgs branch with the states on the Coulomb branch leaves the intrinsic cohomology as states on the Higgs branch that have no counterpart on the Coulomb branch. For this reason the physical states associated to the intrinsic cohomology were given the name {\it pure-Higgs} states in \cite{deBoer:2012ma}. Their number, or more precisely a refined index ${\mathscr B}(\k_0,\k_1,N_1;t)$ counting them, can be obtained by subtracting from the 'total' refined index $\cali^-(\k_0,\k_1,N_1;t)$ the induced part \eqref{indiscoul}, as in \eqref{refindsplit}.
 
 As discussed in the introduction it is conjectured that pure-Higgs states exist only in the scaling regime. Translating the scaling condition \eqref{scalreg} into the geometric parameters \eqref{parmap1} reproduces exactly the geometric existence criterion of section \ref{exist}, which proves the conjecture for $2+1$ quivers. Interestingly the argument of section \ref{exist} does not depend on the detailed geometric/cohomological structure of the Grassmannian, but appears mainly sensitive to the $\mathbb{CP}$ factor.
 
 We finish our discussion of the negative chamber with an explicit formula for $\cali^{-}(\k_0,\k_1,N_1;t)$, from which ${\mathscr B}(\k_0,\k_1,N_1;t)$ can be computed. The starting point is \eqref{indspec}. Using well known expressions for the characteristic classes relevant to the case of interest (\ref{negamb}, \ref{negbund}), see e.g. \cite{Borel}, one finds (see the appendix for some techniques in working on the Grassmannian)
 \begin{align}
 \cali^- (\k_0,\k_1,N_1; t) =& \frac{t^{N_1^2-\k_0+1}}{N_1!} \oint \frac{d y}{1- (1-t^2)\,y} \,  \prod_{a=1}^{N_1} d\, z_a \left\{\left(\frac{1+t^2\,y}{y}\right)^{\k_0}  \prod_{a=1}^{N_1} \left(\frac{1+z_a}{z_a} \right)^{\kappa_1} \prod_{a\neq b} (z_a - z_b) \right. \nonumber \\
 &\qquad \qquad \qquad \qquad \left.\prod_{a\,,\, b} \left( \frac{1}{(1+z_a) - t^2 \, z_b} \right) \;  \prod_{a=1}^{N_1} \left( \frac{y+ z_a}{1+ t^2\, y+ z_a} \right)^{\kappa_1} \right\}\;, \label{3nodes-threshold-ref-index-2}
 \end{align}
 Here we re-expressed $h$, (minus) the Chern-class of the hyperplane bundle of $\mathbb{CP}^{\kappa_0-1}$ as $e^{-h} = 1-(1-t^2) \, y$. Similarly $x_i$, the Chern roots of the tautological bundle of $\Gr(N_1,\kappa_1)$, were rewritten as $e^{-x_i} = \frac{1}{1+(1-t^2)\, z_i}$.
 This result here obtained using the algebraic topology of $\calv^-$ as the zero locus of a vector bundle section,  matches a computation using the abelianization method of \cite{Lee:2013yka}.

Using \eqref{3nodes-threshold-ref-index-2} and (\ref{indiscoul}, \ref{Cgen2}) one can now compute the pure-Higgs index ${\mathscr B}({\k_0,\k_1,N_1; t})$, we provide examples and discuss some properties in section \ref{secex}.

\paragraph{The positive chamber}In this chamber the FI parameters satisfy $\theta_{0}>0$ and $\theta_1>0$, in that case the only choice consistent with the first three equations of \eqref{HiggsBranch} is $\varphi_{0\bar0}{}^{k}=0$. The equations then simplify to
\begin{eqnarray}
  \sum_{k=1}^{\k_1}\bar\varphi_{\bar 01}{}^{ka}\varphi_{\bar 01}{}^k{}_b&=&\theta_1 \delta^a_b +\sum_{k=1}^{\k_1} \varphi_{01}{}^{ka}\bar \varphi_{01}{}^k{}_b\label{poseq1}\\
\sum_{k=1}^{\kappa_1}\sum_{a=1}^{N_1}\bar \varphi_{01}{}^k{}_a\varphi_{01}{}^{ka}&=&\theta_0 \label{poseq2}\\
\sum_{k_1=1}^{\kappa_\m}\sum_{k_2=1}^{\kappa_\mu}\sum_{a=1}^{N_1}\, C_{1 k_1 k k_2}\varphi_{0 1}{}^{k_1 a}\varphi_{\bar 0 1}{}^{k_2}{}_a&=&0\label{poseq3}
\end{eqnarray}
Again we see that the first two equations can be solved independently to provide an ambient space $\calm$, while the third equation restricts to a subspace inside $\calm$. Equation \eqref{poseq2} defines, upon the U(1) quotient, a $\mathbb{CP}^{\kappa_1N_1-1}$. Contrary to the negative chamber equation \eqref{poseq1} is not fully decoupled. But observe that the right hand side is a positive definite Hermitian matrix and can thus be diagonalized.  Defining $M^a{}_b=\theta_1 \delta^a_b +\sum_{k=1}^{\k_1} \varphi_{01}{}^{ka}\bar \varphi_{01}{}^k{}_b$ we can thus write $M=UDU^\dagger$. Let then $T=UD^{1/2}$, in terms of the new variables
\begin{equation}
\bar\phi_{\bar 01}{}^{ka}=\sum_{b=1}^{N_1}\left(T^{-1}(\varphi_{01},\bar \varphi_{01})\right)^a{}_b\bar\varphi_{\bar 01}{}^{kb}
\end{equation}
equation \eqref{poseq1} then takes the form
\begin{equation}
\sum_{k=1}^{\k_1}\bar\phi_{\bar 01}{}^{ka}\phi_{\bar 01}{}^k{}_b=\delta^a_b
\end{equation}
Upon the U($N_1$) quotient the solutions to this equation define the Grassmannian $\Gr(N_1,\kappa_1)$ and so we can conclude that in the positive chamber
\begin{equation}
\calm_+=\mathbb{CP}^{\kappa_1N_1-1}\times\Gr(N_1,\kappa_1)\label{posamb}
\end{equation}
In the new variables the equations \eqref{poseq3} read
\begin{equation}
\sum_{k_1=1}^{\kappa_\m}\sum_{k_2=1}^{\kappa_\mu}\sum_{a,b=1}^{N_1}\, C_{1 k_1 k k_2}\phi_{\bar 0 1}{}^{k_2}{}_b \left(T^{\dagger}(\varphi_{01},\bar \varphi_{01})\right)^b{}_a\varphi_{0 1}{}^{k_1 a}=0\label{poseq4}
\end{equation} 
Similarly to the negative chamber we would like to interpret the equations above as the vanishing of a section of a vector bundle $\cale_+$ over $\calm_+$. Since the free index $k$ runs from $1$ to $\kappa_0$, we expect this vector bundle to have a trivial $\mathbb{C}^{\k_0}$ factor in addition to another factor of complex dimension 1, a line bundle $\call$:
\begin{equation}
\cale_+=\mathbb{C}^{\kappa_0}\otimes \call \qquad \mathrm{rank}_\mathbb{C}\call=1\qquad \call=?\label{posbund}
\end{equation}
This picture is only partially satisfactory, since we could not precisely identify this line bundle and furthermore it appears it is not holomorphic. Nonetheless we will assume the Higgs branch $\calv_+$ to have the structure of a vanishing locus of a vector bundle section and the Lefschetz-Sommese theorem to apply. As we will see the results obtained under this assumption are consistent with those obtained in the negative chamber. To use some of the results of section \ref{LSsec} we can identify the parameters as
\begin{equation}
a=\kappa_1N_1-1\qquad d_0=N_1(\kappa_1-N_1)\qquad r=\kappa_0\label{parmap2}
\end{equation}
The induced part of the cohomology of $\calv_+$ is sensitive only to the rank of the vector bundle, using the map between Grassmannian cohomology and partitions, as in the negative chamber, and the parameters above one finds that also in the positive chamber
\begin{equation}
\cali_{\mathrm{ind}}^+(\kappa_0,\kappa_1,N_1;t)=\cali_{\mathrm{C}}^+(\kappa_0,\kappa_1,N_1;t)   \label{pos-cham-ind-coul-eql}
\end{equation}
where looking up the right hand side in \eqref{coulind} one should remember that in the positive chamber $\kappa_\theta=2\kappa_1N_1-\kappa_0$. So also in this chamber, under the assumptions we made, the induced cohomology on the Higgs branch can be identified with the states on the Coulomb branch, leaving the intrinsic cohomology to be identified with the pure-Higgs states.

\begin{figure}
	\centering
	\includegraphics[scale=1]{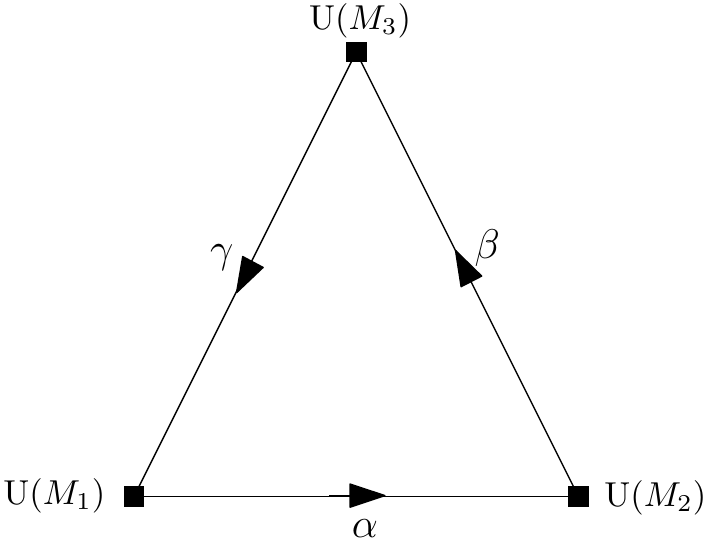}
	\caption{Parameterization of a generic 3 node quiver.\label{3quivex}}
\end{figure}

To find the index ${\mathscr B}({\kappa_0,\kappa_1,N_1};t)$ we need to find the total refined index. Since we were not able to fully identify the relevant vector bundle we cannot use the algebraic topology techniques we used in the negative chamber to compute the refined index. Instead we resort to the abelianization method of \cite{Lee:2013yka}. The idea underlying this method is to translate the calculation on the D-term vacuum manifold to a larger toric manifold, of which the D-term equations define a subspace. Let us consider a three node quiver with gauge groups $U(M_i)$, $i=1,2,3$ connected by a number of arrows that are respectively $\alpha, \beta $ and $\gamma$ and that form a loop, as depicted in figure \ref{3quivex}. In the presence of a generic super potential one set of bifundamental scalars will need to vanish, choosing them to be those associated to the arrows from node 3 to 1, the method provides the following formula for the refined index of the Higgs branch:  
\begin{align}
\cali (\mm\,;\,t) =& \frac{t^{-\dim_\mathbb{C}\mm}\left(1-t^2\right)^{1-M_1-M_2-M_3}}{M_!! \, M_2! \, M_3 !} \,\int_{\widetilde{\calm}} \, \prod_{a=1}^{M_1} \prod_{b=1}^{M_2} \left[\frac{(- J_a + K_b) \, \left(1- t^2 \, e^{ J_a - K_b} \right)}{1-e^{J_a - K_b}} \right]^\a \, \, \, \nonumber \\
& \qquad \prod_{b=1}^{M_2} \prod_{c=1}^{M_3} \left[ \frac{(- K_b + L_c)\,\left(1-t^2 \, e^{K_b - L_c} \right)}{1- e^{K_b - L_c}} \right]^{M_2} \, \prod_{a=1}^{M_1} \, \prod_{c=1}^{M_3} \left( \frac{1-e^{ J_a - L_c}}{1-t^2\, e^{ J_a - L_c}} \right)^\g \, \nonumber \\
& \qquad \prod_{a \neq a'} \left(\frac{1 - e^{J_a - J_{a'}}}{1 - t^2\, e^{ J_a - J_{a'}}}\right) \, \prod_{b \neq b'} \left(\frac{1-e^{K_b - K_{b'}}}{1- t^2 \, e^{K_b - K_{b'}}}\right) \, \prod_{c\neq c'} \left(\frac{1- e^{L_c-L_{c'}}}{1-t^2 \, e^{L_c-L_{c'}}} \right) \;, \label{tot-ref-ind-numb-Ab-prop}
\end{align}
where the integration is over the toric quiver variety $\widetilde{\calm}$ and one needs to put one of the 2-forms in the set $\{J_i\}  \cup\{K_j\}\cup \{L_k\}$ to zero, to decouple the overall U(1) gauge group.

For the 2+1 quiver in the chamber at hand the parameters can be identified as follows
$$ M_1=M_3 =1 \quad, \quad M_2 = N_1 \quad, \quad \a=\beta=\k_1 \quad, \quad \g = \k_0 \;. $$
We thus have the 2-forms $J, K_a, L$. Choosing  $L$ to vanish, formula \eqref{tot-ref-ind-numb-Ab-prop} takes the form
\begin{align*}
\cali^+(\k_0,\k_0,N_1\,;\,t) =& \frac{t^{1+\k_0+N_1^2-2N_1\kappa_1}}{N_1! \left(1-t^2\right)^{N_1+1}} \,\int_{\widetilde{\calm}} \, \prod_{a=1}^{N_1} \left[\frac{( J - K_a) \, \left(1- t^2 \, e^{ -J + K_a} \right)}{1-e^{-J+ K_a}} \right]^{\k_1} \left( \frac{1-e^{ -J}}{1-t^2\, e^{-J}} \right)^{\k_0}   \\
& \qquad\qquad\qquad\qquad \prod_{a=1}^{N_1} \left[ \frac{( K_a)\,\left(1-t^2 \, e^{-K_a } \right)}{1- e^{-K_a}} \right]^{\k_1} \, \, \prod_{a \neq b} \left(\frac{1-e^{K_a - K_{b}}}{1- t^2 \, e^{K_a - K_{b}}}\right) \,  \;, 
\end{align*}
The integration of the product of 2-forms over the manifold $\widetilde{\calm}$ can be rewritten as a contour integral around the origin:
\begin{equation}
J,K_a\quad\leftrightarrow\quad h,x_a\qquad \mbox{and}\qquad \int_{\widetilde{\calm}}\quad\leftrightarrow\quad \oint dh \prod_{a=1}^{N_1}\frac{dx_a}{(h-x_a)^{\k_1} x_a^{\k_1}}
\end{equation}
with the prescription\footnote{Note that this prescription can equivalently be interpreted as taking $|z_i|<|h|$ and thus the $h$ contour surrounding all $z_i$. 	} that one should first perform the integral over all $x_i$, before the integral over $h$.

After an additional change of variables $ e^{-h} = \frac{1}{1+(1-t^2)\,y} \;,\ e^{-x_i} = \frac{1}{1+(1-t^2)\,z_i} \;, $ we get
\begin{align}
\cali^+ (\kappa_0,\kappa_1,N_1\;;\;t) = \frac{t^{1+\k_0+N_1^2-2N_1\kappa_1}}{N_1 !} &\oint \frac{d y}{1+ (1-t^2) \, y} \, \prod_{a=1}^{N_1} dz_a \,\left\{ \left(\frac{y}{1+y} \right)^{\k_0} \,\prod_{a=1}^{N_1} \left( \frac{1+z_a}{z_a} \right)^{\k_1} \right.\nonumber \\
&\left. \prod_{a\neq b} (z_a - z_b)\,\prod_{a\,,\, b} \left( \frac{1}{1+z_a - t^2 \, z_b} \right)  \, \prod_{a=1}^{N_1} \left(\frac{1+y- t^2\, z_a}{y-z_a} \right)^{\k_1} \right\} \;. \label{3nodes-scaling-ref-index-2}
\end{align}

Although both the induced index $\cali^+_{\mathrm{ind}}$ and the total index $\cali^+$ differ from their counterparts in the negative chamber their difference ${\mathscr B}$, the index enumerating intrinsic cohomology/pure-Higgs states is independent of the chamber it is computed in. To see why this is the case, note that upon changing the integration variable $y\rightarrow -(1+t^2y)$ in \eqref{3nodes-scaling-ref-index-2} the integrand becomes exactly minus that of \eqref{3nodes-threshold-ref-index-2}. The only, but key, difference with \eqref{3nodes-threshold-ref-index-2} is that by making the change of variables the contour integral appearing in $\cali^+$ is now around $y=-t^{-2}$ instead of around $y=0$ as for $\cali^-$. It follows\footnote{Notice that to make the comparison to section \ref{LSsec} we first rewrite the contour integrals over $z_a$ as geometric integrals over corresponding 2-forms $z_a$. Then we can freely change the order of integration without worrying about contour prescriptions. It should be possible to rederive the same result using a change of contour prescription, but we believe our method is practically the most efficient.} that the difference $\Delta\cali=\cali^--\cali^+$ is exactly of the form \eqref{deltotind}. Similarly, by comparing \eqref{parmap2} to \eqref{parmap1}, we see that the Higgs branch in the positive chamber is related to that of the negative chamber by the change $a\rightarrow r-1$, $r\rightarrow a+1$, so that $\Delta\cali_\mathrm{ind}=\cali^-_\mathrm{ind}-\cali^+_\mathrm{ind}$ is of the form \eqref{dinddef}. By the factorization equality of section \ref{facsec} it thus follows that
\begin{eqnarray}
{\mathscr B}(\kappa_0,\kappa_1,N_1;t)&=&\cali^+(\kappa_0,\kappa_1,N_1;t)-\cali_\mathrm{ind}^+(\kappa_0,\kappa_1,N_1;t)\label{pH1}\\
&=&\cali^-(\kappa_0,\kappa_1,N_1;t)-\cali_\mathrm{ind}^-(\kappa_0,\kappa_1,N_1;t)\label{pH2}
\end{eqnarray}
This has the important interpretation that the pure-Higgs states are stable under wall-crossing.  

\subsubsection{Pure-Higgs states: examples and symmetries}\label{secex}
The information on the pure-Higgs states is collected in the index ${\mathscr B}$ defined in \eqref{bdef} and given by the precise formula (\ref{pH1}, \ref{pH2}) in case of the 2+1 quiver. We collected the non-vanishing indices for the first few values of $\kappa_1$ in table \ref{pHtable}.

\begin{table}[p]
	\centering
	\begin{tabular}{|c|c|}
		\hline
		$\kappa_1$ & $N_1$\\
		\hline 
		$\kappa_0$ & $\hat{\mathscr B}(\k_0,\k_1,N_1;u)$\\
		\hline
	\end{tabular} \\
	
	\vspace{1cm}
	
	\begin{tabular}{|c|c|}
		\hline
		$2$ & $1$\\
		\hline 
		$2$ & $1$\\
		\hline
	\end{tabular}\qquad\qquad\quad\ \begin{tabular}{|c|cc|}
	\hline
	3 & 1 & 2\\
	\hline 
	2 & 2 &\\
	3 & $u$&\\
	4 &5&\\
	5 && 5$^{(*)}$\\
	6 && $u$\\
	7 && 2\\
	\hline
\end{tabular}\qquad\qquad\quad\ \begin{tabular}{|c|ccc|}
\hline
4 & 1 & 2 & 3\\
\hline 
2 & 3& &\\
3 & $3u$& &\\
4 &$u+18$& &\\
5 &$11u$& 19 &\\
6 &19&$21u$ &\\
7 &&$7u^2+58$ &\\
8 &&$u^3+33u$ &\\
9 &&$7u^2+58$ &\\
10 &&$21u$ &19\\
11&&19 &$11u$\\
12&& &$u+18$\\
13 && &$3u$\\
14 && &3\\
\hline
\end{tabular}

\vspace{1cm}

\begin{tabular}{|c|cccc|}
	\hline
	5 & 1 & 2 & 3 & 4\\
	\hline 
	2&	4 &  &  &  \\
	3&	$6 u$ &  &  &  \\
	4&	$4u^2+43$ &  &  &  \\
	5&	$u^3+51u$ & 49 &  &  \\
	6&	$19 u^2+128$ & $101 u$ &  &  \\
	7&	$71 u$ & $84 u^2+398$ &  &  \\
	8&	69 & $36u^3+476u$ &  &  \\
	9&	 & $9 u^4+313u^2+1141$ &  &  \\
	10&	 & $u^5+101 u^3+1003u$ & 174 &  \\
	11&	 & $14 u^4+508 u^2+1701$ & $351 u$ &  \\
	12&	 & $91 u^3+1026u$ & $264u^2+1073$ &  \\
	13&	 & $264u^2+1073$ & $91 u^3+1026u$ &  \\
	14&	 & $351u$ & $14 u^4+508 u^2+1701$ &  \\
	15&	 & 174 & $u^5+101 u^3+1003u$ &  \\
	16&	 &  & $9 u^4+313u^2+1141$ &  \\
	17&	 &  & $36u^3+476u$ & 69 \\
	18&	 &  & $84 u^2+398$ & $71u$ \\
	19&	 &  & $101u$ & $19u^2+128$ \\
	20&	 &  & 49 & $u^3+51u$ \\
	21&	 &  &  & $4 u^2+43$ \\
	22&	 &  &  & $6u$ \\
	23&	 &  &  & 4 \\
	\hline
\end{tabular}
\caption{Pure-Higgs indices of $2+1$ quivers, for the first few values of $\kappa_1$, in transformed form, see \eqref{uform}. Only non-zero values are shown, the indices exist for all other choices of $\kappa_0$ and $N_1$ as well, but vanish there. {}$^{(*)}$This particular case already appeared in \cite{Lee:2013yka} eqn (3.18).\label{pHtable}}
\end{table}

As pointed out in the previous subsections, two important properties of this index are the independence of the FI parameters and the vanishing outside the scaling regime. In addition we observe the following symmetries:
\begin{eqnarray}
{\mathscr B}(\k_0,\k_1,N_1;t)&=&{\mathscr B}(\k_0,\k_1,N_1;t^{-1})\label{sym1}\\
 {\mathscr B}(\k_0,\k_1,N_1;t)&=&(-1)^{\k_0+N_1+1}{\mathscr B}(\k_0,\k_1,N_1;-t)\label{sym2}\\
{\mathscr B}(\k_0,\k_1,N_1;t)&=&{\mathscr B}(\kappa_1^2-\k_0,\k_1,\kappa_1-N_1;t)\label{sym3}
\end{eqnarray}
The first of this symmetries is equivalent to the statement that the number intrinsic cohomology classes is symmetric in $p$ and $q$: $\beta^{(p,q)}=\beta^{(q,p)}$. The pure-Higgs index inherits this property directly from $\cali$ and $\cali_{\mathrm{ind}}$, which are each invariant under $t\rightarrow t^{-1}$. The second property follows directly from the definition \eqref{bdef}. Together the symmetries (\ref{sym1}, \ref{sym2}) imply that ${\mathscr B}(t)$ is a Laurent polynomial in $t$ that is invariant under $t\rightarrow t^{-1}$, and that is either even or odd in $t$, depending on its degree. It follows that we can equivalently re-express the index as a polynomial $\hat{\mathscr B}(u)$ with positive integer coefficients via\footnote{More formally this is an integral transform: $\hat{\mathscr B}(u)=\oint \frac{ {\mathscr B}(t)}{1+ut}\frac{dt}{t}$. The inverse is simply ${\mathscr B}(t)=\hat{\mathscr B}(-t)+\hat{\mathscr B}(-t^{-1})-\hat{\mathscr B}(0) $.}
\begin{equation}
{\mathscr B}(t)=b_0+\sum_{n=1}^{n_\mathrm{max}}(-1)^nb_n(t^n+t^{-n})\qquad \hat{\mathscr B}(u)=\sum_{n=0}^{n_\mathrm{max}}b_n u^n\label{uform}
\end{equation}
As $\hat{\mathscr B}(u)$ collects the same information more compactly we have chosen to use this form in table \ref{pHtable}.

The third symmetry \eqref{sym3} is somewhat surprising and more non-trivial. It has its origin in the factor $\Gr(N_1,\kappa_1)$ in the ambient space, see (\ref{posamb}, \ref{negamb}), and the canonical duality $\Gr(N_1,\kappa_1)\cong \Gr(\kappa_1-N_1,\kappa_1)$. This symmetry can be observed in the examples of table \ref{pHtable} and we provide a derivation for the general case in appendix \ref{app-grassm-stuff}.

\subsection{Arbitrary $n$}
Using the experience on the $n=1$ case, we now generalize our methods to the generic case where $n$ is an arbitrary positive integer. We will be very brief in the following, restricting ourselves mainly to stating results. This is because the results derived in the $n=1$ case generalize in a straightforward manner to the present case. The reason for this is that these $n$ non-abelian nodes hardly 'interact' with each other due to our restriction on the arrows in the quiver. The only mild interaction goes through the two U(1) nodes $0$ and $\overline{0}$ to which all non-abelian nodes are connected, see figure \ref{2nquiv}. This can be seen at the level of equations as well, see equations \eqref{Cbranch} and \eqref{HiggsBranch}. Roughly, we expect that the modification that one will find in studying this generic case, be it the Coulomb or Higgs branch, has two parts. A trivial part that is essentially duplicating the description associated to the U$(N_1)$ node, leading to the appearance of the index $\mu$ taking values $1,2,\ldots,n$. And a less trivial part that involves a slight modification to the description associated to the two U(1) nodes.   

As in the previous subsection we will start our discussion with the Coulomb branch before  getting to the Higgs branch. As already mentioned in the beginning of this section we choose the FI parameters so that we have only two distinct chambers: the positive and negative chamber.     
\subsubsection{Coulomb branch states}
As we did in the $n=1$ case, we want to evaluate the function
\begin{equation}
\cali_\mathrm{C}(\kappa_0, \{\kappa_\m, N_\m\};t)=\sum_{\mbox{\scriptsize{states on CB}}} t^{2m^\mathrm{tot}} \;,\label{coulind-generic}
\end{equation}
where $m^\mathrm{tot}$ denote the $J_3^{\mathrm{total}}$ eigenvalue of a state. Following the discussion in subsection \ref{section-n1-coulmb}, the total angular momentum in the present case is the sum of the angular momentum $\vec{J}^0$ of the two U(1) plus the angular momentum of particles associated to the nodes U$(N_\m)$. Hence, it is straightforward to conclude  that the description of angular momentum states given by equations \eqref{dyonic-ang-mom-descipt} and \eqref{defjk} read in the present case (see \cite{deBoer:2009un} for details):
\begin{align}
0 \leq m^1_\m < m^2_\m < \ldots < m^{N_\m}_\m < \kappa_\m  \;, \qquad -j^\mathrm{tot}\leq m^\mathrm{tot}\leq j^\mathrm{tot}
\end{align}
where
\begin{equation}
j^\mathrm{tot}= \frac{\k_\theta-\sum_\m N_\m-1}{2} - \sum_\m \sum_{a=1}^{N_\m} m^a_\m \quad\mbox{and}\quad \kappa_\theta=\begin{cases}
\kappa_0 &\; \theta \in \mbox{negative chamber}\\
2\sum_\m \kappa_\m N_\m-\kappa_0 &\; \theta \in \mbox{positive chamber}
\end{cases}\label{defjk-gen}
\end{equation}
This simple generalization allows us to mimic the steps used in the $n=1$ case to construct the generating function:
\begin{equation}
\mathbb{I}_\mathrm{C}(\{\kappa_\m,N_\m\}; t,z)=\sum_{\kappa_\theta=0}^\infty \cali_\mathrm{C}(\kappa_0,\{\kappa_\m,N_\m\};t) \; z^{\kappa_\theta} \;.
\end{equation}
We find:
\begin{equation}
\mathbb{I}_\mathrm{C}(\{\kappa_\m,N_\m\}; t,z)=\frac{z^{\sum_\m N_\m^2+1}}{(1-tz)(1-t^{-1}z)}\; \prod_\m \begin{bmatrix}\kappa_\m\\N_\m\end{bmatrix}_{z^2}
\end{equation}
This allows us to evaluate the refined index $\cali_\mathrm{C}(\kappa_0,\{\kappa_\m,N_\m\};t)$ as the contour integration:
\begin{equation}
\cali_\mathrm{C}(\kappa_0,\{\kappa_\m,N_\m\};t)=\oint dz\frac{z^{\sum_\m N_\m^2-\kappa_\theta}}{(1-tz)(1-t^{-1}z)} \; \prod_\m \begin{bmatrix}\kappa_\m\\N_\m\end{bmatrix}_{z^2}\label{Cgen1-gen}
\end{equation}

We close this subsection by mentioning the criterion for the existence of BPS states in the Coulomb branch. We have: 
\begin{quote}
	{\bf Coulomb existence criterion}: The Coulomb branch is empty\\
	 {}\qquad \hspace{10cm}a) in the negative chamber when $\kappa_0 \leq \sum_\m N_\m^2$,\\
	 \qquad b) in the positive chamber when $\kappa_0 \geq \sum_\m N_\m (2\kappa_\m-N_\m)$.
\end{quote}
The regime of parameters where both chambers exist correspond to the scaling regime. The latter is the regime where spin 0 multiplets appear in the spectrum of BPS states. In the present case, we find:
\begin{quote}
{\bf Scaling regime}: The Coulomb branch supports a spin 0 multiplet iff
\begin{equation}
\sum_\m N_\m^2<\kappa_0< \sum_\m N_\m (2\kappa_\m-N_\m)\label{scalreg-gen}
\end{equation}
\end{quote}
As before, this regime depends on the 'charge' parameters only and is independent of the FI terms. Let us now turn to the Higgs branch.

\subsubsection{Higgs branch states} \label{arb-n-higgs-sect}
The discussion here follows closely subsection \ref{sect-Higgs-n1}. As was done there, we need to study each chamber on its own. The final results turn out to be a rather straightforward generalization of the $n=1$ case. 
\paragraph{The negative chamber:}
In this chamber we have $\theta_0<0$ and $\theta_\m>0$ for all $n$ nodes. This choice of FI terms corresponds to the vanishing of all the fields $\varphi_{0\m}{}^{ka}$. Hence, equations 
\eqref{HiggsBranch} reduce to
\begin{eqnarray}
\sum_{k=1}^{\k_\m} \bar\varphi_{\bar 0\m}{}^{ka}\varphi_{\bar 0\m}{}^k{}_b&=&\theta_\m \delta^a_b \label{HBneg1-gen}\\
\sum_{k=1}^{\k_0}\bar \varphi_{0\bar 0}{}^k \varphi_{0\bar 0}{}^{k}&=&-\theta_0 \label{HBneg2-gen} \\
\sum_{k_1=1}^{\kappa_0}\sum_{k_2=1}^{\kappa_\m}\, C_{\m k k_1 k_2}\varphi_{0\bar 0}{}^{k_1}\varphi_{\bar 0 \m}{}^{k_2}{}_a&=&0\label{HBneg3-gen}
\end{eqnarray}
Note that there are $n$ equations in the first and last lines. It is clear that a decoupled structure similar to the one in the $n=1$ is present here as well. To see this, compare these equations to \eqref{HBneg1}, \eqref{HBneg2}, and \eqref{HBneg3}. Following the same strategy as before, the first two lines define the ambient manifold:
\begin{equation}
\calm_-=\mathbb{CP}^{\kappa_0-1}\times \prod_\m \Gr(N_\m,\kappa_\m)\,.\label{negamb-gen}
\end{equation}
The last line then defines the vanishing locus of a section of the vector bundle:
\begin{equation}
\cale_-= \oplus_\m \left( H^*\otimes S_\m^*\otimes\mathbb{C}^{\kappa_\m} \right) \label{negbund-gen}
\end{equation}
where $H^*$ is the dual of the hyperplane bundle of $\mCP^{\k_0 -1}$, $S_\m^*$ is the dual of the tautological bundle of $\Gr(N_\m,\kappa_\m)$, and $\mbbc^{\k_\m}$ is the trivial vector bundle of rank $\k_\m$. Hence, geometrically the Higgs branch in this chamber can be identified with $\calv_-$, the vanishing locus of a section of the vector bundle $\cale_-$ over the ambient manifold $\calm_-$. This puts us into the geometric setting of section \ref{LSsec} as before. For example, the cohomology of $\calv_-$ splits into an induced part and an intrinsic part. The study of them will be the subject of the remaining of this subsection. We will just state the results since the technical part is similar to the one in subsection \ref{sect-Higgs-n1}.

The first important result is that the induced cohomology is isomorphic to the Coulomb branch states. Actually, the equality \eqref{indiscoul} generalizes to the present case:
 \begin{equation}
 \cali_{\mathrm{ind}}^-(\kappa_0,\{\kappa_\m,N_\m\};t)=\cali_{\mathrm{C}}^{-}(\kappa_0,\{\kappa_\m,N_\m\};t)\label{indiscoul-gen}
 \end{equation}
 with the right hand side given in \eqref{Cgen1-gen} with $\kappa_\theta=\kappa_0$. As explained in section \ref{LSsec}, the refined index of the intrinsic cohomology can be computed by taking the difference between the total and induced indices. This brings us to next point which is finding an expression for the total refined index $\cali^- (\kappa_0,\{\kappa_\m,N_\m\};t)$. Following the same strategy as in the $n=1$ case, one easily finds that: 
 \begin{align}
 \cali^-(\kappa_0,\{\kappa_\m,N_\m\};t) = t^{1-\k_0} \; \oint \frac{d y}{1- (1-t^2)\,y} \; \left\{ \left(\frac{1+t^2\,y}{y}\right)^{\k_0} \; \prod_\m I^-_\m (\kappa_\m,N_\m\,;\,h\,;\,t) \right\}   \label{3nodes-threshold-ref-index-gen}
 \end{align}
 where: 
 \begin{align}
 I^-_\m (\kappa_\m,N_\m\,;\,h\,;\,t) =&\frac{t^{ N_\m^2}}{N_\m!} \oint \prod_{a=1}^{N_\m} \frac{d\, z_a}{1+(1-t^2) \, z_a} \, \left\{ \prod_{a=1}^{N_\m} \left(\frac{1+z_a}{z_a} \right)^{\kappa_\m} \prod_{a\neq b} \left( \frac{z_a - z_b}{(1+z_a) - t^2 \, z_b} \right) \; \right. \nonumber \\
 &\qquad \qquad \qquad \qquad\qquad \qquad\qquad \left.  \prod_{a=1}^{N_\m} \left( \frac{y+ z_a}{1+ t^2\, y+ z_a} \right)^{\kappa_\m} \right\}\;, 
 \end{align}
Let us now turn to the positive chamber.
\paragraph{The positive chamber:}
Once again we will be very brief in the following restricting ourselves to the results most of the time. This is because most of the technical details and arguments generalize in a straightforward manner from the $n=1$ case. The choice of FI parameters is such that $\q_0>0$ and $\q_\m>0$ for all the nodes $U (N_\m)$. This choice corresponds to setting all the fields $ \varphi_{0\bar0}{}^{k}=0 $. Hence, equations \eqref{HiggsBranch} become:
\begin{align}
  &\sum_{k=1}^{\k_\m}\bar\varphi_{\bar 0\m}{}^{ka}\varphi_{\bar 0\m}{}^k{}_b = \theta_\m \delta^a_b +\sum_{k=1}^{\k_\m} \varphi_{0\m}{}^{ka}\bar \varphi_{0\m}{}^k{}_b\label{poseq1-gen}\\
&\sum_{\m=1}^n \sum_{k=1}^{\kappa_\m}\sum_{a=1}^{N_\m}\bar \varphi_{0\m}{}^k{}_a\varphi_{0\m}{}^{ka} = \theta_0 \label{poseq2-gen}\\
&\sum_{\m=1}^n \sum_{k_1=1}^{\kappa_\m}\sum_{k_2=1}^{\kappa_\mu}\sum_{a=1}^{N_\m}\, C_{\m k_1 k k_2}\varphi_{0 \m}{}^{k_1 a}\varphi_{\bar 0 \m}{}^{k_2}{}_a=0\label{poseq3-gen}
\end{align}
Following the same steps as in the $n=1$, it is easy to see that the ambient manifold takes the form:
\begin{equation}
\calm_+=\mathbb{CP}^{\sum_\m \kappa_\m N_\m -1}\times \prod_{\m=1}^n \Gr(N_\m,\kappa_\m)\label{posamb-gen}
\end{equation}
Unfortunately, for the same reasons as before, we do not have a clear understanding of this chamber as a vanishing locus of a vector bundle over $\calm_+$. Despite this, we will assume that the Lefschetz-Sommese theorem is applicable to this case as well. Under this assumption, it is easy to figure out that equation \eqref{pos-cham-ind-coul-eql} remains valid:
\begin{equation}
\cali_{\mathrm{ind}}^+(\kappa_0,\kappa_1,N_1;t)=\cali_{\mathrm{C}}^+(\kappa_0,\kappa_1,N_1;t) \label{pos-cham-ind-coul-eql-gen}
\end{equation}
where the right hand side is defined by equation \eqref{Cgen1-gen} for $\kappa_\theta=2\sum_\m \kappa_\m N_\m-\kappa_0$. This means that the intrinsic cohomology is identified with the pure-Higgs states. To find their index ${\mathscr B}(\kappa_0,\{\kappa_\m,N_\m\};t)$ we need to find the total refined index. Using the Abelianization method of \cite{Lee:2013yka} and following similar steps as before, the equation \eqref{3nodes-scaling-ref-index-2} generalizes to:
\begin{align}
 \cali^+ (\kappa_0,\{\kappa_\m,N_\m\}\;;\;t) = t^{1+\k_0} \; \oint \; \frac{d y}{1+ (1-t^2) \, y} \; \left(\frac{y}{1+y} \right)^{\k_0} \; \prod_\m I^+ (\kappa_\m,N_\m\;;\;t\;;\;y) \;, \label{3nodes-scaling-ref-index-2-gen}
\end{align}
where:
\begin{align*}
 I^+ (\kappa_\m,N_\m\;;\;t\;;\;y) =&  \frac{t^{N_\m^2-2N_\m \kappa_\m}}{N_\m !} \; \oint \prod_{a=1}^{N_\m} dz_a \; \left\{\left( \frac{1+z_a}{z_a} \right)^{\k_1} \; \prod_{a\neq b} (z_a - z_b) \right.\\
&\qquad \qquad \qquad \qquad \left. \,\prod_{a\,,\, b} \left( \frac{1}{1+z_a - t^2 \, z_b} \right)  \, \prod_{a=1}^{N_1} \left(\frac{1+y- t^2\, z_a}{y-z_a} \right)^{\k_1}  \right\} \;.
\end{align*}
The contour integration over the $z_a$'s is around the origin, hence one works in the regime $|z_a|<h$. As a result one gets negative powers of $h$ in the expression of $I^+ (\kappa_\m,N_\m\;;\;t\;;\;y)$ which allows to perform the contour integration of $h$ around the origin in the expression of $\cali^+ (\kappa_0,\{\kappa_\m,N_\m\}\;;\;t)$. We close this subsection by discussing some properties of the pure-Higgs states.

\subsubsection{Pure-Higgs states}
We will only collect here the results since they are just a straightforward application of results in section \ref{LSsec} upon association of the following parameters:
\begin{align}
 \textrm{Negative Chamber:} \qquad  &a = \k_0 -1 \quad ,\quad d_0 = \sum_\m N_\m \, (\k_\m - N_\m) \quad,\quad r = \sum_\m N_\m \, \k_\m   \label{gen-map-sect-2-neg} \\
 \textrm{Positive Chamber:} \qquad  &a = \sum_\m N_\m \, \k_\m -1 \quad ,\quad d_0 = \sum_\m N_\m \, (\k_\m - N_\m) \quad,\quad r = \k_0   \label{gen-map-sect-2-neg}
\end{align}
The following results then follow:

\begin{quote}{\bf Existence criterion:} The pure-Higgs states exist if and only if:
\begin{equation}
\sum_\m N_\m^2<\kappa_0< \sum_\m N_\m (2\kappa_\m-N_\m)\label{pHiggs-ex-gen}
\end{equation}
\end{quote}
This is the same regime of parameters where both chambers are non-empty. This is a confirmation of conjecture 3) of the introduction.

\begin{quote}
{\bf Stability under wall crossing:} The pure-Higgs index states ${\mathscr B}(\kappa_0,\{\kappa_\m,N_\m\};t)$ is the same in both positive and negative chambers.
\end{quote}{}
As in the $n=1$ case this follows because the results of section \ref{facsec} both the complete refined index and its induced part can be shown to have the same wall-crossing jump, so that their difference ${\mathscr B}(\kappa_0,\{\kappa_\m,N_\m\};t)$ is invariant. This validates conjecture 2) of the introduction.

We refrain from providing explicit examples in the generic case, since there are simply too many parameters to vary. But of course the reader can easily explore ${\mathscr B}(\kappa_0,\{\kappa_\m,N_\m\};t)$ by computing the difference between \eqref{3nodes-threshold-ref-index-gen} and \eqref{Cgen1-gen}, at least for some small values of the parameters. Let us point out that the ${\mathscr B} (\kappa_0,\{\kappa_\m,N_\m\};t)$ retains the same three properties (\ref{sym1}, \ref{sym2}, \ref{sym3}) of the $n=1$ case. In particular the third property becomes:
\begin{equation}
 {\mathscr B} (\k_0,\{\k_\m,N_\m\}; t) = {\mathscr B} \left(\sum_\m \k_\m^2-\k_0,\{\k_\m,\k_\m - N_\m\}; t\right)  \;.
\end{equation}

\section*{Acknowledgements}
We thank D. Mirfendereski for an initial collaboration on this project and B. Pioline for making the Mathematica package \texttt{CoulombHiggs.m} freely and publicly available and for kindly answering our questions. 

IM and DVdB were supported by TUBITAK grant 113F164 during the first part of this research. DVdB is partially supported by TUBITAK grant 117F376. IM is partially supported by DGRSDT.

\appendix
\section{Some Properties of complex Grassmannians} \label{app-grassm-stuff}
In this appendix we will summarize those properties of complex Grassmannians $\mGr (m\,,\, N)$ that are of most relevance to this paper. For more details see e.g \cite{Griffiths:433962, hirzebruch1966topological, Borel}. Throughout this appendix, $\mGr (m\,,\, N)$ stands for the Grassmannian of $m$-planes in $\mbbc^N$. Our exposition will be very brief as proofs will be omitted.  

We start by reviewing the cohomology of complex Grassmannians. Then, we briefly discuss their Chern character and Todd genus. After that, we explain how to rewrite the integration over a Grassmannian as a contour integration. Finally we prove the symmetry property \eqref{sym3}, which is essentially due to the canonical duality $\mGr (m, N)\cong \mGr (N-m, N)$.
\subsection{The cohomology of complex Grassmannians}
The only non-trivial cohomology of $\mGr (m\,,\, N)$ is of even degree. The associated Betti numbers $b^{(2k)}$ correspond to the number of ways to partition $k$ into $m$ parts each of which is smaller or equal $(N-m)$. This implies for example that the Poincar\'e polynomial of $\mGr (m\,,\, N)$ is given by:
\begin{equation}
 \calp_{\mGr (m \,,\, N)} (q) = \begin{bmatrix}N\\m\end{bmatrix}_{q^2} \;. \label{grass-poincr-polynom}
\end{equation}

Hence, we can associate to each non-trivial element of the cohomology of $\mGr (m\,,\, N)$ a Young diagram $\l$ that sits inside a box of with $m$ rows and $(N-m)$ columns. This bijective map:
$$ \l \, \longleftrightarrow \O_\l \,,$$
where $\O_\l$ is a non-trivial element of the cohomology of $\mGr (m\,,\,N)$, extends to a ring homomorphism between the ring cohomology of $\mGr (m\,,\, N)$ and the ring of Schur polynomials $\cals_\l (x)$ of $m$ variables, with $l$ as described earlier. This homomorphism allows us to map any calculation on the cohomology of $\mGr (m\,,\,N)$ to one on Schur polynomials.  
\subsection{Characteristic classes of complex Grassmannians}
Remember that the points of the space $\mGr (m\,,\,N)$ are parameterized by $m$ orthogonal vectors $v_i$ in $\mbbc^N$. Our starting point in constructing the characteristic classes of $\mGr (m\,,\,N)$ is the exact sequence:
\begin{equation}
 0 \, \lra \, S \, \lra \, V \, \lra \, Q \, \lra \, 0 \;, \label{grass-defn-short-sequence}
\end{equation}
where $S$, the tautological vector bundle, is the vector bundle whose fiber is the $m$ dimensional vector space generated by the vectors $v_i$. The fiber of $V$ is the trivial $\mbbc^N$ vector space, and the fiber of $Q$ is the perpendicular complement of the fiber of $S$ inside $\mbbc^N$. The tangent space of $\mGr (m\,,\,N)$ is then isomorphic to:
\begin{equation}
  \calt \left( \mGr (m\,,\,N) \right) \cong S^\ast \otimes Q  \;, \label{grss-tang-space}
\end{equation}
where $S^\ast$ is the dual of $S$. This relation together with the short exact sequence \eqref{grass-defn-short-sequence} are enough to construct the multiplicative characteristic classes of $\mGr (m\,,\,N)$. This can be achieved using the $m$ Chern roots $\{x_i\}$ of $S^\ast$ or the $n$ Chern roots $\{y_j\}$ of $Q$. We will show how that works for the multiplicative class $ \ttc_{(m\,,\,N)} (t) $ of tangent bundle of $\mGr (m\,,\,N)$ defined by equation \eqref{new-mult-class}. We have from equation \eqref{grss-tang-space}:
\begin{align}
\ttc_{(m\,,\,N)} = \prod_{i=1}^m \prod_{j=1}^{N-m} \left( \frac{(x_i+y_j) \,\left( 1 - t \, e^{-(x_i+y_j)} \right) }{1 - e^{-(x_i+y_j)}} \right) \;. \label{grass-proto-charac-class}
\end{align}
Using now the short exact sequence \eqref{grass-defn-short-sequence} we find that:
\begin{align}
 \prod_{i=1}^m \left( 1 - t \, e^{\mp x_i} \right) \; \prod_{j=1}^{N-m} \left( 1 - t \, e^{\pm y_j} \right) &= (1-t)^N \;, \label{ch-chrac-reln}\\
 \prod_{i=1}^m \left( 1\pm t \, x_i \right) \; \prod_{j=1}^{N-m} \left( 1 \mp t \, y_j\right) &= 1 \;. \label{todd-genus-reln}
\end{align}
The first equation above is a result of the properties of the Chern character, whereas the second one is associated to the total Chern class. Using these constraints together with equations \eqref{grass-proto-charac-class} we find at the end:
\begin{align}
 \ttc_{(m\,,\, N)} (t) &= \left( \frac{1}{1-t} \right)^m \prod_{i=1}^m \left[ \frac{x_i \, \left( 1-t \, e^{-x_i} \right)}{1- e^{-x_i}} \right]^N \prod_{k \neq \ell} \left[ \frac{1 - e^{x_\ell -x_k} }{(x_\ell - x_k) \, \left( 1 - t \, e^{x_\ell - x_k} \right)} \right] \;, \\
\ttc_{(m\,,\, N)} (t) &= \left( \frac{1}{1-t} \right)^{N-m} \prod_{j=1}^{N-m} \left[ \frac{y_j \, \left( 1-t \, e^{-y_j} \right)}{1- e^{-y_j}} \right]^N \prod_{k \neq \ell} \left[ \frac{1 - e^{y_\ell -y_k} }{(y_\ell - y_k) \, \left( 1 - t \, e^{y_\ell - y_k} \right)} \right]  \;.
\end{align}
Next, Let us look for the meaning of the integration over a Grassmannian $\mGr (m\,,\,N)$ when we map the calculation to Schur polynomials. 
\subsection{Integration over complex Grassmannians}
Relying on the ring homomorphism we mentioned in the beginning of this appendix, we can re-express integration over $\mGr (m\,,\,N)$ as a contour integration over $m$ variables. The idea is as follows. The integration over the Grassmannian $\mGr (m\,,\,N)$ is equivalent to picking the coefficient in front of the element of the top cohomology $H^{2(N\,m-m^2)}$. This is associated to the Schur polynomial:
$$ \cals_{top} (x) = \prod_{i=1}^m x_i^{N-m} \;, $$
that is associated to the Young diagram $\l$ which is a box with $m$ rows and $N-m$ columns. Hence, the integration over $\mGr (m\,,\,N)$ is the same as choosing the coefficient in front of the aforementioned Schur polynomial when we rewrite the polyform as a symmetric function in the $\{x_i\}$. In the present case, using the properties of Schur polynomials (see e.g. \cite{macdonald1998symmetric}), one easily finds that:
\begin{equation}
  \int_{\mGr (m\,,\,N)} F (x) = \frac{1}{m!} \oint \left(\prod_{i=1}^m \frac{d x_i}{x_i^N }\right) \; F(x) \; \left(\prod_{k \neq \ell} (x_k -x_\ell)\right)   \;, \label{grass-intgrn-origin}
\end{equation}
where on the right hand side, the contour integration is around the origin. We can also use the Chern roots of the vector bundle $Q$ instead. In this case the integration formula reads:
\begin{equation}
  \int_{\mGr (m\,,\,N)} F (y) = \frac{1}{(N-m)!} \oint \left(\prod_{j=1}^{N-m} \frac{d y_j}{y_j^N }\right) \; F(y) \; \left(\prod_{k \neq \ell} (y_k -y_\ell)\right)   \;. \label{grass-intgrn-dual}
\end{equation}
Let us discuss the symmetry that such dual expressions imply for the total refined index and pure-Higgs states.
\subsection{A Symmetry for the pure-Higgs states}
As mentioned in the section \ref{LSsec}, the way we calculate the refined index of the pure-Higgs states is by taking the difference between the total and induced refined indices of the chamber under consideration. Hence, in the following we will deal with all of these refined indices.

Following the discussion in section \ref{sect-3nodes-hggs-branch}, the total refined index of the negative chamber is given by: 
\begin{align*}
 \cali (\k_0,\k_1,N_1; t) &= \frac{t^{1+N_1^2-\k_0}}{N_1 !} \left( \frac{1}{1-t^2} \right)^{1+N_1} \oint d h \, \prod_{i=1}^{N_1} dx_i\,\left\{ \prod_{k \neq \ell} \left( \frac{1- e^{x_k -x_\ell}}{1-t^2\, e^{x_k -x_\ell}} \right)  \, \; \right. \nonumber \\
                            & \left. \left(\frac{1- t^2 \, e^{-h}}{1- e^{-h}} \right)^{\k_0} \prod_{j=1}^{N_1} \left( \frac{1- t^2 \, e^{-x_j}}{1- e^{-x_j}} \right)^{\k_1}  \, \prod_{i=1}^{N_1} \left( \frac{1 - e^{- (h +x_i)}}{1 - t^2 \, e^{-(x_i + h)}} \right)^{\k_1}\right\} \;. 
\end{align*}
Let us make the following change of variables:
$$ u = t\, \left(\frac{1-e^{-h}}{1-t^2\, e^{-h}}\right) \, \Lra \, e^{-h} = \frac{1-t^{-1} \,u}{1-t \, u} \;. $$
Then, the expression above becomes:
\begin{align*}
 &\cali (\k_0,\k_1,N_1; t) = \frac{t^{N_1^2}}{N_1 !} \left( \frac{1}{1-t^2} \right)^{N_1} \oint \frac{d u}{u^{\k_0}} \, \prod_{i=1}^{N_1} dx_i\,\left\{ \prod_{k \neq \ell} \left( \frac{1- e^{x_k -x_\ell}}{1-t^2\, e^{x_k -x_\ell}} \right)  \, \; \right. \nonumber \\
            &\qquad \qquad \left. \frac{1}{(1-t^{-1} \, u)(1-t \, u)} \;\prod_{j=1}^{N_1} \left( \frac{1- t^2 \, e^{-x_j}}{1- e^{-x_j}} \right)^{\k_1}  \, \prod_{i=1}^{N_1} \left( \frac{1-t\, u - (1- t^{-1} \,u) \,e^{-x_i}}{1-t\, u - t^2 \, (1 -t^{-1} \, u) \, e^{-x_i }} \right)^{\k_1}\right\} \;. 
\end{align*}
This implies that the generating function:
\begin{equation}
 \mbi (\k_1,N_1; t;q) = \sum_{\k_0=0}^\infty \cali (\k_0,\k_1,N_1; t) \; q^{\k_0} \;,  \label{part-egen-funct-tot-ref-indx-def}
\end{equation}
takes the form:
\begin{align}
 \mbi (\k_1,N_1; t;q) = \frac{q}{(1-t^{-1} \, q) (1-t \, q)} \;  \mbi_0 (\k_1,N_1; t;q) \;, \label{part-egen-funct-tot-ref-indx-exprn-1}
\end{align}
where the function $\mbi_0 (\k_1,N_1; t;q)$ is defined as follows: 
\begin{align}
 \mbi_0 (\k_1,N_1; t;q) &= \frac{t^{N_1^2}}{N_1 !} \left( \frac{1}{1-t^2} \right)^{N_1} \oint \prod_{i=1}^{N_1} dx_i\,\left\{ \prod_{k \neq \ell} \left( \frac{1- e^{x_k -x_\ell}}{1-t^2\, e^{x_k -x_\ell}} \right)  \, \; \right. \nonumber \\
                            & \qquad \qquad \left. \prod_{j=1}^{N_1} \left( \frac{1- t^2 \, e^{-x_j}}{1- e^{-x_j}} \right)^{\k_1}  \, \prod_{i=1}^{N_1} \left( \frac{1-t\, q - (1- t^{-1} \,q) \,e^{-x_i}}{1-t\, q - t^2 \, (1 -t^{-1} \, q) \, e^{-x_i }} \right)^{\k_1}\right\} \;. \label{part-egen-funct-tot-ref-indx-core-exprn-1}
\end{align}
We want to re-express this generating function using the dual variables $y_j$. Using the constraint \eqref{ch-chrac-reln}, we find:
\begin{align*}
 \prod_{i=1}^{N_1} \left( \frac{1-t\, q - (1- t^{-1} \,q) \,e^{-x_i}}{1-t\, q - t^2 \, (1 -t^{-1} \, q) \, e^{-x_i }} \right) = t^{\k_1 - 2 N_1} \; q^{\k_1} \; \prod_{j=1}^{\k_1-N_1} \left( \frac{1-t\, q^{-1} - (1- t^{-1} \,q^{-1}) \,e^{-y_j}}{1-t\, q^{-1} - t^2 \, (1 -t^{-1} \, q^{-1}) \, e^{-y_j }} \right) \;.
\end{align*}
Hence, we get the following relation:
\begin{align}
  \mbi_0 (\k_1,N_1; t;q) = q^{\k_1^2} \; \mbi_0 (\k_1,\k_1-N_1; t;q^{-1})   \;. \label{imp-prop-1}
\end{align}
Before dealing with the induced generating function, let us first rewrite the expression \eqref{part-egen-funct-tot-ref-indx-core-exprn-1} is a more tractable form. To do that we start by making the following change of variables:
$$ e^{-x_i} = \frac{1}{1+ (1-t^2) \, z_i} \; \Lra \; z_i = \frac{e^{x_i}-1}{1-t^2} \;.  $$
After some straightforward manipulations, we find:
\begin{align}
 \mbi_0 (\k_1,N_1; t;q) &= \frac{t^{N_1^2}}{N_1 !} \, \oint \prod_{i=1}^{N_1} \frac{d z_i}{z_i^{\k_1}}\,\left\{ \prod_{k \neq \ell} \left( \frac{ z_k - z_\ell}{1+z_k-t^2\,z_\ell} \right)  \, \prod_{i=1}^{N_1} \frac{1}{1+(1-t^2) \, z_i} \; \right. \nonumber \\
                            & \qquad \qquad \qquad\qquad \left. \prod_{j=1}^{N_1} \left( 1+z_j \right)^{\k_1}  \, \prod_{i=1}^{N_1} \left( \frac{t^{-1}\, q + (1- t \,q) \,z_i}{1+(1 -t \, q) \, z_i} \right)^{\k_1}\right\} \;. \label{part-egen-funct-tot-ref-indx-core-exprn-2}
\end{align}
This expression allows us to show that:
\begin{align*}
 \mbi_0 (\k_1,N_1; q;q) &= q^{N_1^2} \; \calp_{\mGr (N_1 \,,\, \k_1)} (q)\;, \\
 \mbi_0 (\k_1,N_1; q^{-1};q) &= q^{N_1^2} \; \calp_{\mGr (N_1 \,,\, \k_1)} (q) \;,
\end{align*}
where $\calp_{\mGr (N_1 \,,\, \k_1)} (q)$ is the Poincar\'e polynomial of the Grassmannian $\mGr (N_1 \,,\, \k_1)$. The two equations above imply that:
\begin{equation}
 \mbi_0 (\k_1,N_1; t;q) = q^{N_1^2} \; \calp_{\mGr (N_1 \,,\, \k_1)} (q) + (1-t^{-1} \, q) (1-t \, q) \; \d \, \mbi_0 (\k_1,N_1; t;q) \;. \label{tot-cohom-mater-eqn}
\end{equation}

Let us now switch gears and talk about the induced cohomology. From equations \eqref{indiscoul}, \eqref{coulomb-refined-indx-gen-funct-def}, and \eqref{coulomb-refined-indx-gen-funct-exprn}, its generating function reads:
\begin{equation}
 \mbi_{ind} (\k_1,N_1; t;q) = \frac{q^{1+N_1^2}}{(1-t q) (1-t^{-1} q)} \; \calp_{\mGr (N_1 \,,\, \k_1)} (q) \;.
\end{equation}
Hence, the generating function associated to the intrinsic cohomology reads:
\begin{equation}
 \mbb (\k_1,N_1; t;q) = \sum_{\k=0}^\infty {\mathscr B} (\k_0,\k_1,N_1; t) \; q^{\k_0}  = q \; \d \, \mbi_0 (\k_1,N_1; t;q) \;. 
\end{equation}
From the property \eqref{imp-prop-1} and expression \eqref{tot-cohom-mater-eqn}, one finds that:
\begin{equation}
 \mbb (\k_1,N_1; t;q) = q^{\k_1^2} \;\; \mbb (\k_1,\k_1-N_1; t;q^{-1}) \;.
\end{equation}
Since $\k_0$ the powers of $q$ that appear in the generating functions are positive by construction, then  this identity implies that:
\begin{equation}
 \k_0 \leq \k_1^2 \;.    \label{n-scaling-mild-constrnt}
\end{equation}
This is in agreement with the scaling condition. Next, from the definition of the generating function $\mbi (\k_1,N_1; t;q)$ given by equation \eqref{part-egen-funct-tot-ref-indx-def}, one concludes that:
\begin{align}
 {\mathscr B} (\k_0,\k_1,N_1; t) = {\mathscr B} (\k_1^2-\k_0,\k_1,\k_1-N_1; t)  \;,  \label{Dieter-scnd-id}
\end{align}
which is the symmetry we set out to prove.

We close this appendix by discussing this symmetry in the case of $2+n$ node quivers. It is easy to see, following the discussion of the negative chamber in subsection \ref{arb-n-higgs-sect}, that a similar albeit more involved symmetry still exists. Following the same steps as before one finds that: 
\begin{align}
 {\mathscr B} (\k_0,\{\k_\m,N_\m\}; t) = {\mathscr B} \left(\sum_\m \k_\m^2-\k_0,\{\k_\m,\k_\m - N_\m\}; t\right)  \;.  \label{Dieter-scnd-id-gen}
\end{align}

\bibliographystyle{JHEP}
\bibliography{refslist}

\end{document}